\documentclass[11pt]{article}
\usepackage{authblk}
\usepackage{graphicx}
\usepackage[margin=1.25in]{geometry}
\usepackage[usenames,dvipsnames]{color}
\usepackage{url}
\usepackage{verbatim}
\usepackage[section]{placeins}
\usepackage{subfig}
\usepackage{scalerel,amssymb}

\usepackage[colorlinks = true,
            linkcolor = blue,
            urlcolor  = blue,
            citecolor = blue,
            anchorcolor = blue]{hyperref}

\usepackage{cleveref}

\newcommand\pubnumber{arXiv Preprint}
\newcommand\pubdate{\today}
\newcommand{\minus}{\scalebox{0.75}[1.0]{$-$}}

\def\Title#1{\begin{center} {\LARGE #1 } \end{center}}
\def\Author#1{\begin{center}{ \sc #1} \end{center}}
\def\Address#1{\begin{center}{ \it #1} \end{center}}

\newcommand\pubblock{\rightline{\begin{tabular}{l} \pubnumber\\
         \pubdate \end{tabular}}}
\newenvironment{Abstract}{\begin{quotation} \begin{center}
                       ABSTRACT
     \end{center}\bigskip  }{\end{quotation}}

\def\beq{\begin{equation}}
\def\eeq#1{\label{#1}\end{equation}}
\def\eeqn{\end{equation}}

\newenvironment{Eqnarray}%
   {\arraycolsep 0.14em\begin{eqnarray}}{\end{eqnarray}}
\def\beqa{\begin{Eqnarray}}
\def\eeqa#1{\label{#1}\end{Eqnarray}}
\def\eeqan{\end{Eqnarray}}

\let\bar=\overbar

\def\lsim{\mathrel{\raise.3ex\hbox{$<$\kern-.75em\lower1ex\hbox{$\sim$}}}}
\def\gsim{\mathrel{\raise.3ex\hbox{$>$\kern-.75em\lower1ex\hbox{$\sim$}}}}

\def\del{\partial}
\def\Dslash{\not{\hbox{\kern-4pt $D$}}}
\def\dslash{\not{\hbox{\kern-2pt $\del$}}}
\def\pslash{\not{\hbox{\kern-2pt $p$}}}
\def\ETmiss{\not{\hbox{\kern-4pt $E$}}_T}

\def\Dlr{\mathrel{\raise1.5ex\hbox{$\leftrightarrow$\kern-1em\lower1.5ex\hbox{$D$}}}}

\def\MSB{{\bar{M \kern -2pt S}}}
\def\msb{{\bar{\scriptsize M \kern -1pt S}}}

\def\drb{{\bar{\scriptsize D \kern -1pt R}}}

\newcommand\snowmass{\begin{center}\rule[-0.2in]{\hsize}{0.01in}\\\rule{\hsize}{0.01in}\\
\vskip 0.1in Submitted to the  Proceedings of the US Community Study\\ 
on the Future of Particle Physics (Snowmass 2021)\\ 
\rule{\hsize}{0.01in}\\\rule[+0.2in]{\hsize}{0.01in} \end{center}}

\begin{document}

\pubblock

\Title{Snowmass 2021 White Paper Instrumentation Frontier 05\\ White Paper 1: MPGDs: Recent advances and current R\&D}

\bigskip 

\Author{
K.~Dehmelt$^{1,*,\dagger}$,
M. Della Pietra$^{2,\dagger}$,
H.~Muller$^{3,4,\dagger}$, 
S.~E.~Tzamarias$^{5,\dagger}$,
A.~White$^{6,*,\dagger}$,
S.~White$^{4,7,\dagger}$,
Z.~Zhang$^{19,\dagger}$, 
M. Alviggi$^2$, 
I.~Angelis$^{5}$, 
S.~Aune$^{20}$, 
J.~Bortfeldt$^{4,22}$, 
M.~Bregant$^{17}$, 
F.~Brunbauer$^{4}$, 
M.T.~Camerlingo$^{4,9}$, 
V.~Canale$^2$, 
V.~D'Amico$^9$, 
D.~Desforge$^{20}$, 
C.~Di~Donato$^2$, 
R.~Di~Nardo$^9$, 
G.~Fanourakis$^{28}$, 
K.J.~Floethner$^{4}$, 
M.~Gallinaro$^{18}$, 
F.~Garcia$^{26}$, 
I.~Giomataris$^{20}$, 
K.~Gnanvo$^{12}$, 
T.~Gustavsson$^{20}$, 
R.~Hall-Wilton$^{15}$, 
P.~Iengo$^{4,8}$, 
F.J.~Iguaz$^{20,21}$, 
M.~Iodice$^{25}$, 
D.~Janssens$^{4}$, 
A.~Kallitsopoulou$^{5}$, 
M.~Kebbiri$^{20}$, 
K.~Kordas$^{5}$, 
C.~Lampoudis$^{5}$, 
P.~Legou$^{20}$, 
M.~Lisowska$^{4}$, 
J.~Liu$^{19}$, 
M.~Lupberger$^{3,4}$, 
S.~Malace$^{12}$,
I.~Maniatis$^{5}$, 
I.~Manthos$^{5,27}$, 
Y.~Meng$^{19}$, 
H.~Natal da Luz$^{10}$, 
E.~Oliveri$^{4}$,
G.~Orlandini$^{4}$, 
T.~Papaevangelou$^{20}$, 
K.~Paraschou$^{5}$, 
F.~Petrucci$^9$,
D.~Pfeiffer$^{14,15}$, 
M.~Pomorski$^{20}$, 
S.~Popescu$^{11}$, 
F.~Resnati$^{4}$, 
L.~Ropelewski$^{4}$, 
A.~Rusu$^{13}$, 
D.~Sampsonidis$^{5}$, 
L.~Scharenberg$^{3,4}$, 
T.~Schneider$^{4}$, 
G.~Sekhniaidze$^8$, 
M.~Sessa$^9$, 
M.~Shao$^{19}$, 
L.~Sohl$^{20}$, 
J.~Toledo-Alarcon$^{16}$, 
A.~Tsiamis$^{5}$, 
Y.~Tsipolitis$^{29}$, 
A.~Utrobicic$^{4}$, 
M.~van Stenis$^{4}$, 
R.~Veenhof$^{4,23,24}$, 
X.~Wang$^{19}$,
Y.~Zhou$^{19}$
}
\noindent $^1$Stony Brook University, Stony Brook, NY, USA
\newline $^2$University and INFN Sez. Napoli, Naples, Italy
\newline $^3$Bonn University, Bonn, Germany
\newline $^{4}$CERN, Geneva, Switzerland
\newline $^{5}$Aristotle University of Thessaloniki, Greece
\newline $^6$University of Texas, Arlington, TX, USA
\newline $^{7}$University of Virginia, Charlottesville, VA, USA
\newline $^8$INFN Sez. Napoli, Naples, Italy
\newline $^9$University and INFN Sez. Roma Tre, Rome, Italy
\newline $^{10}$Czech Technical University in Prague, Prague, Czech Republic
\newline $^{11}$Kansas University, USA
\newline $^{12}$JLab, Newport News, VA, USA
\newline $^{13}$SRS Technology, Meyrin, Switzerland
\newline $^{14}$Bicocca University, Milano, Italy
\newline $^{15}$ESS ERIC, Lund, Sweden
\newline $^{16}$Universitat Polit\`ecnica de Val\`encia, Val\`encia, Spain
\newline $^{17}$IFUSP, S\~ao Paulo, Brasil
\newline $^{18}$LIP Lisbon, Portugal
\newline $^{19}$USTC, Hefei, China
\newline $^{20}$CEA Saclay, France
\newline $^{21}$Synchrotron Soleil, Gif-sur-Yvette, France
\newline $^{22}$LMU Munich, Germany
\newline $^{23}$MEPhI Moscow, Russian Federation
\newline $^{24}$Uludag University, Turkey
\newline $^{25}$INFN Sez. Roma Tre, Rome, Italy
\newline $^{26}$Helsinki Institute of Physics, Helsinki, Finland
\newline $^{27}$University of Birmingham, UK
\newline $^{28}$NCSR Demokritos, Athens, Greece
\newline $^{29}$National Technical University of Athens, Greece

\noindent$^*$ White Paper Coordinators
\newline\noindent$\dagger$ Corresponding LOI author

\medskip

\Address{ }

\medskip
\begin{Abstract}
\noindent
This paper will review the origins, development, and examples of new versions of Micro-Pattern Gas Detectors. The goal for MPGD development was the creation of detectors that could cost-effectively cover large areas while offering excellent position and timing resolution, and the ability to operate at high incident particle rates. The early MPGD developments culminated in the formation of the RD51 collaboration which has become the critical organization for the promotion of MPGDs and all aspects of their production, characterization, simulation, and uses in an expanding array of experimental configurations.\newline
For the Snowmass 2021 study, a number of Letters of Interest were received that illustrate ongoing developments and expansion of the use of MPGDs. In this paper, we highlight high precision timing, high rate application, trigger capability expansion of the SRS readout system, and a structure designed for low ion backflow.
\end{Abstract}
\snowmass

\newpage
\tableofcontents

\def\thefootnote{\fnsymbol{footnote}}
\setcounter{footnote}{0}
\newpage
\section{Executive Summary}
This paper will review the origins, development, and examples of new versions of Micro-Pattern Gas Detectors. The goal for MPGD development was the creation of detectors that could cost-effectively cover large areas while offering excellent position and timing resolution, and the ability to operate at high incident particle rates. Significant development time was invested in understanding the optimal manufacturing techniques for MPGDs, in understanding their operation, and mitigation of undesirable effects such as discharges and ion backflow. The early MPGD developments culminated in the formation of the RD51 collaboration which has become the critical organization for the promotion of MPGDs and all aspects of their production, characterization, simulation, and uses in an expanding array of experimental configurations. RD51 serves as the essential repository and sharing point for information and data on MPGDs. The CERN MPGD Workshop has been the source of essential expertise in production methods, mitigation and correction of manufacturing issues, and the development of MPGDs for specific experimental environments. There is by now an impressive array of MPGDs from the first GEM and Micromegas, now seen in a wide variety of applications and configurations, through the more recent ThickGEMs, and microR-Wells with resistive layer(s) to mitigate discharge effects. MPGDs are now also used jointly with other detector elements, for example, with optical readout and in liquid Argon detectors. In parallel with the MPGD detector developments, there has been an important creation of a general electronics system, the Scalable Readout System (SRS). This system has seen widespread use and is of great utility in allowing integration of a variety of frontend electronics into one data acquisition system.

For the Snowmass 2021 study, a number of Letters of Interest were received that illustrate ongoing developments and expansion of the use of MPGDs. In this paper, we highlight high precision timing, high rate application, trigger capability expansion of the SRS readout system, and a structure designed for low ion backflow.

The RD51 PICOSEC collaboration is developing very fast timing detectors using a two-stage system with a Cerenkov radiator producing photons to impact a photocathode. The photo-electrons are then drifted to and amplified in a micromegas layer. Prototypes have shown that timing for single photons can be achieved at the 45ps level and at 15ps for MIPs. Multi-pad detectors are being developed, and studies applying an artificial neural network to the waveform have shown potential for very precise timing – in agreement with simulations. Potential applications of the PICOSEC technique include precise timing in electromagnetic showers, and time of flight systems for the future Electron Ion Collider.

The next contribution describes the development of micromegas detectors with a resistive layer for spark mitigation and small readout pads designed for high rate operation. Several schemes have been tested for implementing the resistive layer. These include resistive pads covering the readout pads, layers of Diamond-Like Carbon (DLC) across the plane of the detector, and hybrid schemes using both approaches. Prototype detectors have been tested using radioactive sources and X-rays in the RD51 GDD laboratory, as well as particle (muon, pion) beams at CERN and PSI. Comparisons have been made of the rate capability of the various prototypes. Results are reported on the gain variations and operability at high rates of $\mathcal{O}$(10 MHz/cm$^{2}$) for pad and DLC prototypes. Energy resolution measured using X-rays shows a better result for the DLC version vs. the pad version, attributable to the more uniform electric field in the former. Spatial resolution was measured using particle beams with again the DLC configuration showing a better results – at the 100 $\mu$m level. Having demonstrated the desired rate and other characteristics of the small prototypes, a larger prototype is under construction which potentially will also include integration of the readout electronics with the detector.

In parallel with the development of new MPGD detector techniques there has been an evolution of the Scalable Readout System (SRS to SRSe) to include realtime trigger functionality, deep trigger pipelines and a generalized frontend link using  the eFEC, extended Frontend Concentrator backend. This will allow for the expanded use of a variety of frontend ASICs, and ability to use realtime triggering with firmware in an FPGA. A range of possible triggers is possible including, for example, hit combinations and energy sums. Ongoing plans for SRSe foresee script development for FPGAs, establishing a set of initial triggers, testing of the first eFECs, and addressing the needs of specific experiments.

The final example of new MPGD development is for new structures that are designed to restrict ion backflow that can result in performance degradation and/or detector damage. Specifically, multi-mesh MPGD structures have been studied which preserves the desirable features of micromegas while controlling ion backflow. Results with double and triple mesh structures, with adjusted and optimized gaps, have shown that high gain and very low ion backflow can be achieved offering the prospect of excellent performance in future detector applications.

The ongoing development of MPGDs, evidenced by the many activities of the RD51 collaboration, and the examples given of new techniques, point the way to expanded use and many beneficial applications of MPGDs in future experiments.

\section{Development of the Micro-Pattern gaseous detector technologies: an overview of the CERN-RD51 collaboration}
The modern photo-lithographic technology on flexible and standard PCB substrates favored the invention, in the last years of the 20$^{th}$ century, of novel Micro-Pattern Gas Detectors (MPGD); among them: the Micro-Strip Gas Chamber (MSGC) \cite{ch02-ref1}, the Gas Electron Multiplier \cite{ch02-ref2} (GEM) and the MicroMegas (MM) \cite{ch02-ref3}. Since the very beginning, the goal was the development of novel detectors with very high spatial $\mathcal{O}(50\,\mu$m) and time (ns) resolution, large dynamic range, high-rate capability (up to 10$^6$ Hz/mm$^2$), large sensitive area and radiation hardness - making them an invaluable tool to confront future detector challenges at the frontiers of research. The dedication of several groups of MPGD developers has led to rapid progress, crowned by new inventions and understanding of the underlying operation mechanisms of the different detector concepts. MPGDs promised to fill a gap between the high-performance but expensive solid-state detectors, and cheap but rate-limited traditional wire chambers. Nevertheless, the integration of MPGDs in large experiments was not rapid, despite of the first large-scale application within the COMPASS experiment \cite{ch02-ref4} at CERN SPS in the early days of the 21$^{th}$ century. In COMPASS, telescopes of MMs and GEM trackers demonstrated reference performance at particle fluxes of 25 kHz/mm$^2$, with space resolution better than 100 $\mu$m and time resolution of  $\mathcal{O}(10$ ns). Thus, the potentiality of MPGD technologies became evident and the interest in their applications has grown in the High Energy Physics (HEP) and Nuclear Physics domains, and beyond. Consequently, it became crucial to consolidate and enlarge the dedicated community to foster further developments and dissemination of MPGD applications in the HEP sector and other fields.

The RD51 CERN-based technological collaboration, started in 2008, now has 90 institutions from 25 countries. Since its foundation, the RD51 collaboration has provided important stimulus, and has become a major focus, for the development of MPGDs. While a number of the MPGD technologies were introduced before RD51 was founded, with more techniques becoming available or affordable, new detection concepts are still being introduced, and existing ones are being substantially improved. The nature and extent of the collaboration activities is reflected in the seven Working Groups (WG), transversal to the RD51 activities, covering all relevant R\&D topics: MPGD technologies and novel structures,  detector characterization, study of the physical phenomena and detector simulations, dedicated electronics tools for read-out and laboratory studies, production and engineering aspects, common test facilities, and dissemination beyond the HEP community - including dedicated education and training. 
Important consolidation of the some better-established MPGD technologies has been reached within the RD51 collaboration, often driven by the working conditions of large collider experiments. One of the breakthroughs came with the development of MM with resistive anodes for discharge mitigation \cite{ch02-ref8}, in the context of the ATLAS New Small Wheel (NSW) project. This concept allows limiting the energy of occasional discharges and results in a protection of both the detector and its front-end electronics and, equally relevant, in a substantial dead-time reduction (time required to re-establish the operational voltage). The resistive anodes have been obtained by a variety of approaches: photo-lithography, screen-printing technologies and sputtering.

The construction of GEM-detectors centers on two main issues: GEM-foil production and preservation of the correct spacing between successive GEMs in multilayer configurations. Initially GEM foils were produced using a double-mask approach with the chemical etching performed from both foil faces. The difficulty of aligning the two masks, limiting the achievable lateral size to 50 cm, led to the development of a single-mask production protocol. It was initially developed for the upgrade of the TOTEM experiment \cite{ch02-ref10}, further used to produce GEM foils for the KLOE2 cylindrical GEM-detector \cite{ch02-ref11} and those of the CMS forward muon spectrometer \cite{ch02-ref12}. The requirement to preserve the constant inter-foil spacing in multilayer GEM detectors of large-size was first fulfilled successfully, though with small dead-zones, by adequate spacers; e.g. in COMPASS \cite{ch02-ref13} and TOTEM \cite{ch02-ref14} trackers. Later, the INFN groups involved in the construction of the GEM-based trackers for LHCb experiment introduced an alternative approach: GEM-foil stretching prior to gluing on support frames \cite{ch02-ref15}. This concept also paved the way to the construction of a cylindrical-GEM detector \cite{ch02-ref11}. Further extension of the stretching technique has been introduced for the CMS forward muon spectrometer \cite{ch02-ref12}: to save a relevant fraction of the assembly time the foils are mechanically fixed onto the frames, where they are mechanically stretched and kept at the correct tension without gluing. This NS2 technique, no-stretch no-spacer method, originally presented by R.~de Olivera at MPGD2013 \cite{ch02-ref16}, has become the basis of the GEM-detector construction for the CMS upgrade. Nowadays, the single-mask GEM technology, together with the NS2 technique, simplifies the fabrication process, resulting in an important minimization of the production time – particularly relevant for large-volume production (Fig. 3). So far, the largest GEM foil production has been finalized and the GEMs are in place for the upgrade of the ALICE Time Projection Chamber (TPC) \cite{ch02-ref17}. The demanding requirements of a TPC, where fine resolution tracking and good dE/dx accuracy are equally relevant, has imposed a detailed and stringent quality assessment protocol. Therefore, this construction effort represents the first fully-engineered large mass-production of MPGDs.

THick GEMs (THGEM), also referred to in the literature as Large Electron Multipliers (LEM), were introduced in parallel by several groups at the beginning of the 21$^{st}$ century \cite{ch02-ref18}. They are derived from the GEM design, scaling up ~10-fold the geometrical parameters and changing the production technology. The Cu-coated polyimide foil of the GEM multipliers is replaced by that of standard PCBs (e.g. FR4) and mechanical drilling produces the holes.

The consolidation of the better-established technologies has been accompanied by the flourishing of novel technologies, often specific to well-defined applications. These technologies are derived from MM and GEM concepts, or they are hybrid approaches combining different MPGDs technologies and gaseous with non-gaseous detectors, or they are entirely new architectures. Novel technologies are illustrated by the following selected examples. The Gridpix counter \cite{ch02-ref20} is obtained applying, by post-processing, a small-scale micromesh directly onto a Timepix chip \cite{ch02-ref21}, therefore obtaining a miniaturized MM with the fine granularity of the Timepix itself. The very fine space resolution has been confirmed using a set of Gridpix units as read-out sensors of the ILC TPC prototype \cite{ch02-ref20}. GEM geometries where extra electrodes are added onto one of the two GEM-foil faces aim at improving the capability of multilayer GEM detectors to trap the ions flowing back from the multiplication region, a feature required for performing gaseous photon detectors and for TPC sensors: Micro Hole and Strip Plates \cite{ch02-ref23} and Cobra \cite{ch02-ref24} architectures have been designed and characterized. 

A promising GEM-derived architecture is that of the $\mu$R-WELL \cite{ch02-ref25}, where the anode is directly placed at the hole bottom, forming the well structure, in order to maximize the collection of the avalanche electrons, and it is realized by a resistive layer for spark protection: in this architecture, a single multiplication layer is present. 

Initially, GEMs had been introduced to act as a preamplifier stage in gaseous detectors \cite{ch02-ref26} and, therefore, the concept of the hybrid approach has been present since the very beginning in the MPGD concepts. More recently, electron multipliers have been coupled to devices capable to detect the luminescent light produced in the amplification process: in GEM TPC with optical read-out \cite{ch02-ref27} and THGEM-based read-out for double-phase liquid Ar detectors \cite{ch02-ref28}. A final MM multiplication stage is added to GEM or THGEM multipliers in order to control the ion backflow making use of the intrinsic ion trapping capability of the MM principle; the GEMs and MM scheme has been proposed  as read-out sensor of the ALICE TPC \cite{ch02-ref17}\footnote{But has not been adapted.}, while THGEMs and MM are the basis structure of novel gaseous photon detectors \cite{ch02-ref30}. A GEM foil facing the Medipix chip \cite{ch02-ref31}, forming the GEMpix detectors \cite{ch02-ref32}, is in use for medical applications as well the treatment of radioactive waste \cite{ch02-ref33}. The $\mu$PIC \cite{ch02-ref34} is a fully industrially produced PCB including anode strips on one face and orthogonal cathode strips on the other face. A regular pattern of uncoated zones is present in the cathode strips; an electric conductor buried in the PCB brings the anode voltage to a dot at the center of each uncoated zones: the multiplication takes place there, thanks to the electric field established between the cathode strips and the anode dot. A resistive coating of the cathode strips ensures the spark tolerance of the detector.

A key development which facilitates and encourages the development and use of MPGDs is the realization of the Scalable Readout System (SRS) \cite{ch02-ref35}. This is a multichannel readout solution for a wide range of detector types and detector complexities. The scalable architecture, achieved using multi-Gbps point-to-point links with no buses involved, allows the user to tailor the system size to his needs. The modular topology enables the integration of different front-end ASICs, giving the user the possibility to use the most appropriate front-end for his purpose or to build a heterogeneous experimental apparatus which integrates different front-ends into the same DAQ system. The SRS system is today available through the CERN store or via the CERN Knowledge Transfer office. More than fifty groups are today using SRS for their R\&D or experiments. The new, extended SRS-e paradigm adds realtime trigger functionality, deep trigger pipelines and a generalized frontend link via the new eFEC concentrator card. Horizontal links synchronize clocks and realtime actions, and the vertical links can be connected as vertical or horizontal readout architectures. Trigger extensions for SRS-e are discussed in section \ref{sec-5} below.

\section{High precision timing with the PICOSEC Micromegas detector}






New challenges in current and future accelerator facilities have set stricter
requirements on the timing and rate capabilities of particle detectors. The
PICOSEC Micromegas detector has proven to time the arrival of Minimum
Ionizing Particles (MIPs) with a sub-25 ps precision. Model predictions and laser beam
tests demonstrated that an optimized PICOSEC design can time single photons
with an accuracy of 45 ps which indicates an improved resolution in timing
MIPs of the order of 15 ps. We propose the implementation of the PICOSEC
detector for timing the arrival of EM showers with very high precision as well
as for Time-of-Flight measurements for particle identification applications.

\subsection{The PICOSEC Micromegas concept}
\label{S:2}

\begin{figure}[ht]
\centering\includegraphics[width=0.7\linewidth]{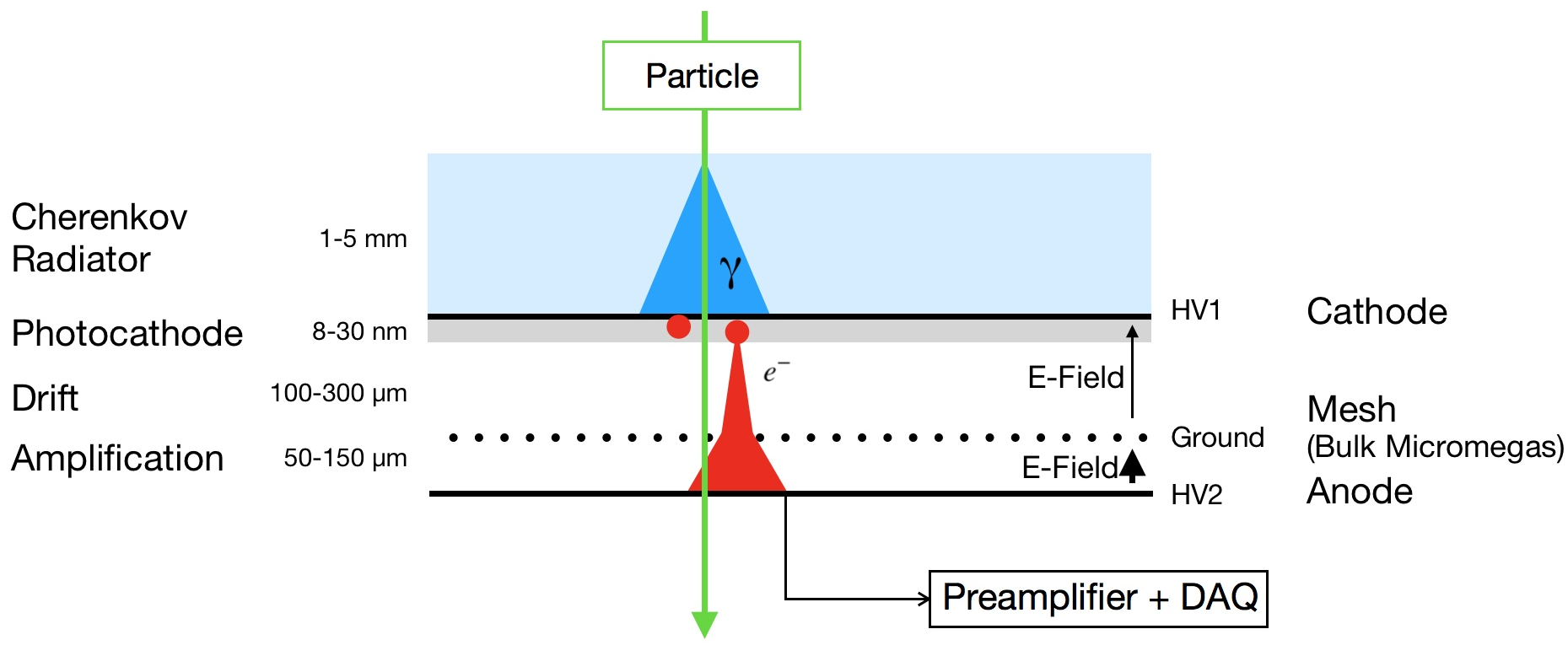}
\caption{Sketch of the PICOSEC-Micromegas working principle.}
\label{fig:concept}
\end{figure}

The PICOSEC detector~\cite{1} consists of a two-stage Micromegas~\cite{2} coupled to a front window that acts as a Cherenkov radiator coated with a photocathode.
A sketch of the detector concept is shown in Fig.~\ref{fig:concept}.
The drift region is very thin ($< 200\mu m$) minimizing the probability of direct gas ionization as well as diffusion effects on the signal timing. Due to the high electric field, photoelectrons undergo pre-amplification in the drift region. The readout is a bulk~\cite{3} Micromegas, which consists of a woven mesh and an anode plane separated by a gap of about 128 mm, mechanically defined by pillars. A relativistic charged particle traversing the radiator produces UV photons, which are simultaneously (RMS less than 10 ps) converted into primary photoelectrons at the photocathode. These primary photoelectrons produce pre-amplification avalanches in the drift region. A fraction of the pre-amplification electrons (about 25$\%$) traverse the mesh and are finally amplified in the amplification region. The arrival of the amplified electrons at the anode produces a fast signal (with a rise-time of about 0.5 ns) referred to as the electron-peak (e-peak), while the movement of the ions produced in the amplification gap generates a slower component for the ion-tail (about 100~ns). This type of detector operated with Neon- or CF$_4$-based gas mixtures can reach high enough gains to detect single photoelectrons.

\subsection{Achievements in precise timing}

The time response of a single-cell PICOSEC prototype detector (circular anode of 1 cm diameter) to single photons and to Minimum Ionizing Particles (MIPs) was extensively studied in laser and particle beams. The PICOSEC waveforms were fully digitized by a fast oscilloscope and analyzed offline to determine the e-peak charge and amplitude as well as the Signal Arrival Time (SAT), which is defined as the time measured at 20$\%$ of the signal amplitude. As reported in Ref.~\cite{1}, the resolution for timing single photons was measured to be 76.0$\pm$0.4 ps, while the accuracy for timing the arrival of muons (MIPs) was measured to be 24.0$\pm$0.3 ps using CsI photocathodes (with an average value of about 11 photoelectrons induced per MIP). Several technologies concerning photocathodes and resistive anodes have been evaluated, which guarantee radiation hardness
and stability in operation while preserving the precise timing capability (28-34 ps) of the detector.
\begin{figure}[ht]
\centering\includegraphics[width=0.9\linewidth]{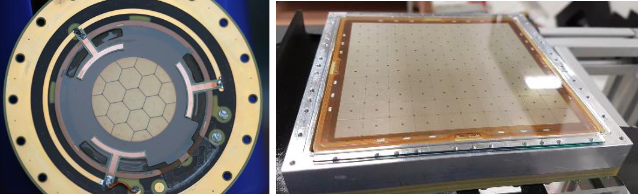}
\caption{Multi-pad PICOSEC prototypes with active areas of about $10cm^2$ (left) and $100cm^2$ (right) respectively.}
\label{fig:MP}
\end{figure}

\begin{figure}[ht]
\centering\includegraphics[width=0.8\linewidth]{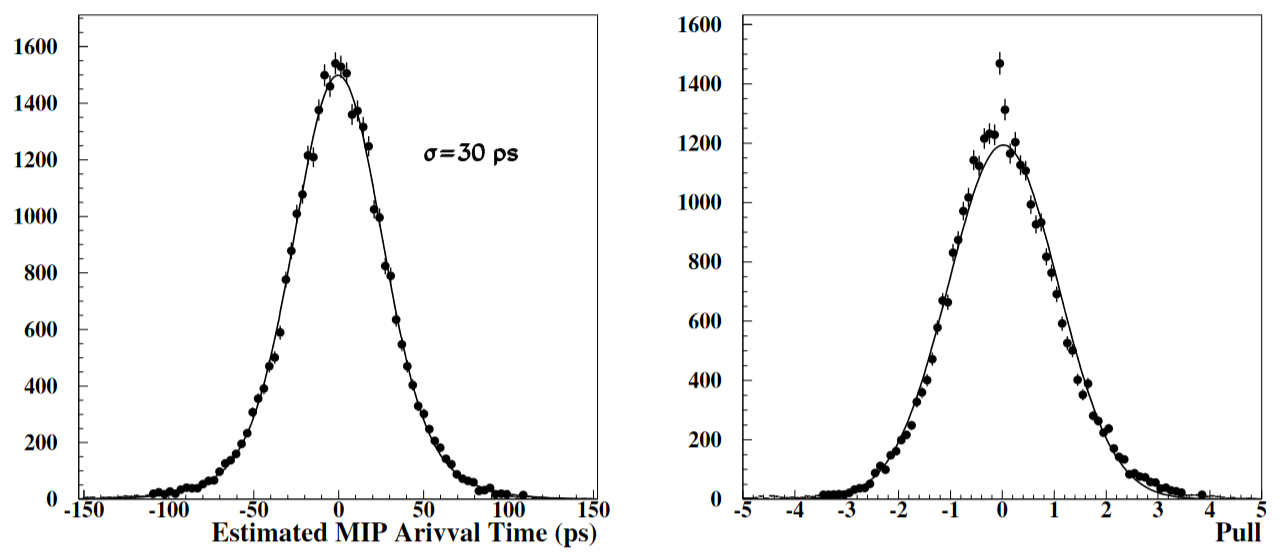}
\caption{(left) Distribution of the signal arrival time of MIPs, passing within 3 mm of a common
corner of four square pads, estimated by combining the individual single-pad
measurements and their expected uncertainties. The solid line represents a fit to the data points
by a sum of two Gaussian functions corresponding to $\sigma$=30~ps. (right) Pull
distribution of estimated signal arrival times. The solid line represents a Gaussian fit to
the data points, consistent with mean and $\sigma$ values equal to 0 and 1, respectively ~\cite{4,sharing}.}
\label{fig:sharing}
\end{figure}

\begin{figure}[ht]
\centering\includegraphics[width=0.7\linewidth]{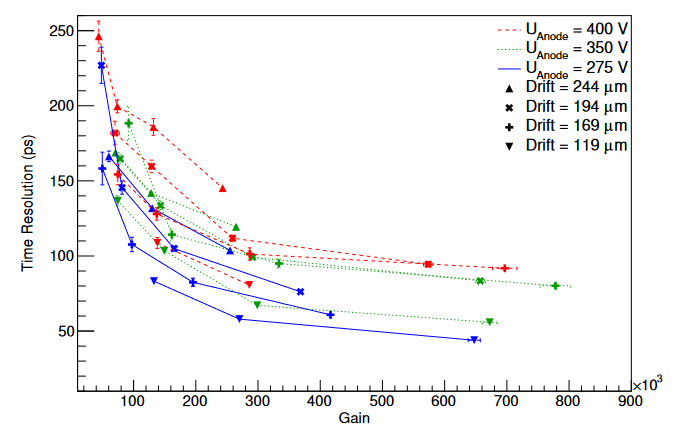}
\caption{Time resolution as a function of the gain for different gap region thicknesses and anode bias
voltages under single photo-electron conditions. The amplification gap is 128 $\mu m$ deep~\cite{8}.}
\label{fig:laser}
\end{figure}

Aiming at large area detectors, multi-pad PICOSEC prototypes were developed, which comprise segmented anodes divided into either hexagonal or square pads (Fig.~\ref{fig:MP}).  
Extensive tests in particle beams revealed that a detector with 19 hexagonal pads, each with 1 cm diameter, offers similar time resolution to the single-pad PICOSEC prototype. This remains true in the case that the incoming MIP induces signals in more than one of the neighboring pads (Fig.~\ref{fig:sharing}). 
Recently, particle-beam tests of an improved, larger prototype (comprising 100  1cm$\times$1cm  pads) demonstrated even better timing capabilities~\cite{10a, 10b,11}. Furthermore, a recent conceptual study~\cite{12}, based on fully digitized PICOSEC waveforms, has shown that existing, cost-effective technologies~\cite{13} can be used for digitizing the signals of large, multi-pad detectors, while retaining the precise timing information. It has also been demonstrated ~\cite{11,12} that a trained Artificial Neural Network (ANN), fed with a partially digitized PICOSEC waveform (e.g. delivered by the SAMPIC~\cite{14} digitizer), provided extremely precise measurement of the signal arrival time. That indicates that PICOSEC detectors could offer timing information in real-time to be utilized for event selection.
The Garfield++ simulation package~\cite{5}, complemented with a custom made simulation of the electronics response, reproduces~\cite{6} very well the PICOSEC timing characteristics. In addition, a phenomenological model~\cite{7} was developed which describes stochastically the dynamics of the signal formation, in excellent agreement with the Garfield++ predictions and the experimental measurements. Recently, guided by the phenomenological model predictions, new PICOSEC designs were advanced that improve significantly the timing performance of the detector. As an example, PICOSEC prototypes with reduced drift gap size (about 120 $\mu m$) reached a resolution of 45 ps (in comparison to 76 ps of the standard PICOSEC prototype) in timing single photons in laser beam tests (Fig.~\ref{fig:laser}).

It should be underlined that such a detector design maintains stable operation when irradiated with intense laser pulses, providing an excellent timing resolution of about  6.8 ps for about 70 photoelectrons. Optimized PICOSEC thin drift-gap design has been recently tested in particle beams and preliminary results~\cite{10a, 10b} confirm the observed performances during laser beam tests~\cite{8} with a resolution below 20 ps for MIPs.

\begin{figure}[ht]
\centering\includegraphics[width=1\linewidth]{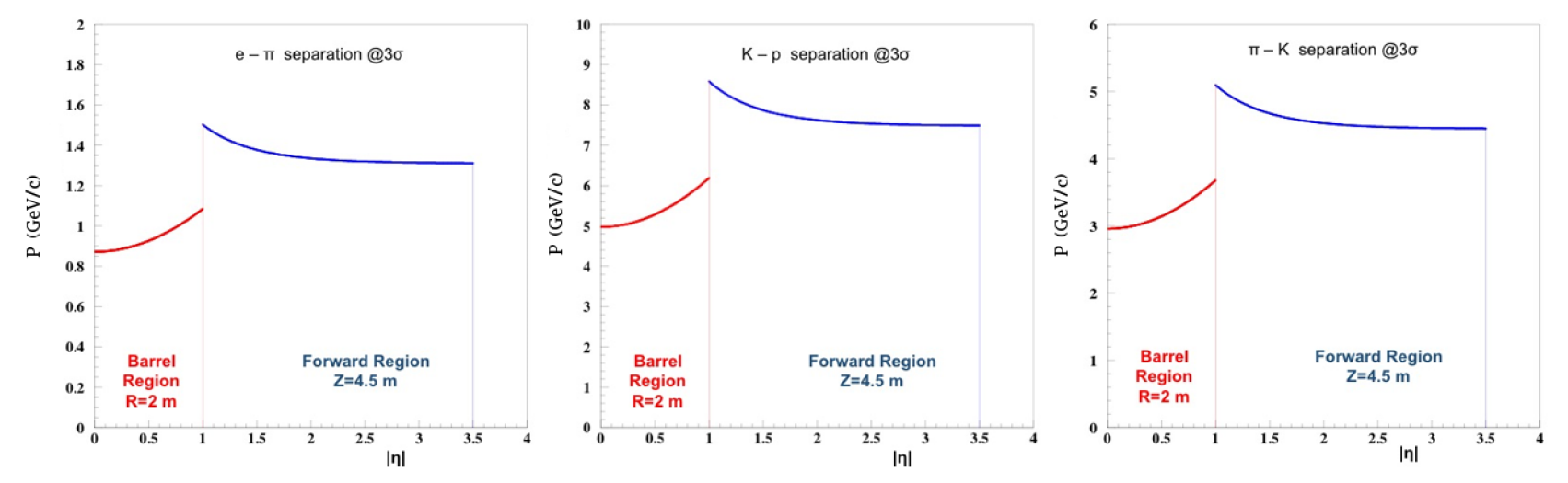}
\caption{Estimated limits (at 3$\sigma$ CL), in the momentum-pseudorapidity plane, for separating $e/\pi$ (left), $K/p$ (center) , and $\pi/K$ (right) in the barrel and forward regions.}
\label{fig:figure}
\end{figure}

\subsection{Potential applications}
The ongoing R\&D in advancing scalable, radiation hard, resistive PICOSEC Micromegas detectors is focused in evaluating new thin-gap Micromegas designs, new photocathode materials, resistive-anode technologies as well as digitizing electronics~\cite{16}. However, at this stage there are detector designs with proven operational virtues offering high timing precision. As an example, a PICOSEC detector embedded in an EM calorimeter could offer an excellent resolution in timing the arrival of EM showers. Indeed, the plethora of secondary relativistic electrons in the EM shower will induce a large number of photoelectrons even in the case that photocathodes (e.g. Al, Cr or Diamond Like Carbon-DLC) with very modest photon-yield are used. Moreover, metallic or DLC photocathodes are almost immune to radiation damage due to the ion back-flow in the Micromegas. According to simulations, a PICOSEC detector can offer a timing resolution of approximately 14 ps for 5 GeV electrons, when the detector is embedded 2 radiation lengths inside the calorimeter. For higher energies, the expected resolution is much less than 10~ps. Tests in electron beams, scheduled for the Spring and Summer of 2022, will  quantify the PICOSEC resolution to time high-energy electromagnetic showers. Recently, a dedicated R\&D has started to design optimized PICOSEC prototypes for the ENUBET (Enhanced Neutrino Beams from Kaon Tagging)~\cite{17} project. These high-precision PICOSEC detectors aim at timing low energy (about 5 GeV) electron showers in the ENUBET  decay tunnel or/and to time-tag electrons from Kaon decays or/and to time muons inside the hadron dump. Another possible application concerns the use of PICOSEC as a TOF detector to provide particle identification at low momenta. We have considered the case of using PICOSEC detectors, with 20 ps timing resolution, to separate e/$\pi$ , $\pi$/K and K/p in the barrel and the forward region of a detector facility at the Electron Ion Collider~\cite{9}. Figure~\ref{fig:figure} shows, in the momentum-pseudorapidity plane,
the separation power at 3$\sigma$ C.L. (e.g.  $3\times\sqrt{2}\times$20~ps) for the barrel and the forward region. Although a time resolution better than 20 ps is needed to reach 3$\sigma$ for $e/\pi$  separation at 4 GeV/c, there is a wide momentum spectrum where the PICOSEC detector offers a good $\pi$/K and K/p separation.

\subsection{Concluding Remarks}

The RD51 PICOSEC collaboration focuses in advancing scalable, radiation hard, resistive PICOSEC Micromegas detectors for very precise timing, by evaluating new thin-gap Micromegas designs, new photocathode materials, resistive-anode technologies as well as digitization electronics. These activities motivate collaboration between groups with diverse expertise, something we are hoping to promote through this letter. However, even at this stage of development, the currently designed detectors
could offer valuable experimental information in physics projects.
It is worth noting that a good timing resolution may also be beneficial to reject out-of-time pileup, with possible use at a future collider facility (FCC, Muon collider, etc.). Another possible application under consideration is to use precise timing at deep underground experiments to differentiate between Cerenkov and scintillation photons where the former provides crucial directional information.





\section{Pixelated resistive MicroMegas for high-rates environment}
The new era of Particle Physics experiments is moving towards new upgrades of present accelerators (Large Hadron Collider at CERN) and the design of high energy 
(tens/hundreds TeV scale) and very high intensity new particle accelerators (FCC-ee/hh, EIC, Muon Collider). Cost effective, high efficiency particle detection in a high 
background and high radiation environment is fundamental to accomplish their physics program. We present a new High Granularity Resistive Micromegas detector capable of ensuring
full efficient and stable operation and an excellent tracking capabilities up to particle fluxes of 10 MHz/cm$^2$.
\subsection{Introduction}
Micro Patterns Gaseous Detectors (MPGDs) will play a crucial role in the future of High Energy Physics, when very high particle intensities will become significant.
Resistive Micromegas (R-MM) are built with parallel plate electrodes structure, with the volume divided into two gaps (drift and amplification) by means of a metallic mesh.
The anode plane hosts the read-out elements, usually strips, built using Printed Circuit techniques. To prevent discharges that could damage the detector and worsen its performance,
a layer of resistive strips facing the amplification gap is added.~\cite{ResMM}. 
This is, for example, the solution adopted by ATLAS for the New Small Wheel upgrade, for operations up to few kHz/cm$^2$~\cite{ATLASNSW}.
Since 2015, our research team has been working on the further development of the MM technology to reach stable and efficient 
operation up to particle fluxes of tens of MHz/cm$^2$, with low occupancy and good stability and robustness, and ensuring good tracking capabilities.
To achieve these levels of performance we designed a higher granularity detector 
using few mm$^2$ large readout pads instead of strips. This choice significantly reduces the occupancy of the readout 
elements, but the resistive structure required for the spark protection needs to be optimized to avoid losing efficiency at very high rates.
In this paper we will report the status of our R\&D project and a summary of the performances measured along last years by mean of high intensity X-ray, muon and pion sources.
\subsection{High granularity MicroMegas for high-rate environments}
\begin{figure}
\begin{center}
\includegraphics[width=0.90\textwidth]{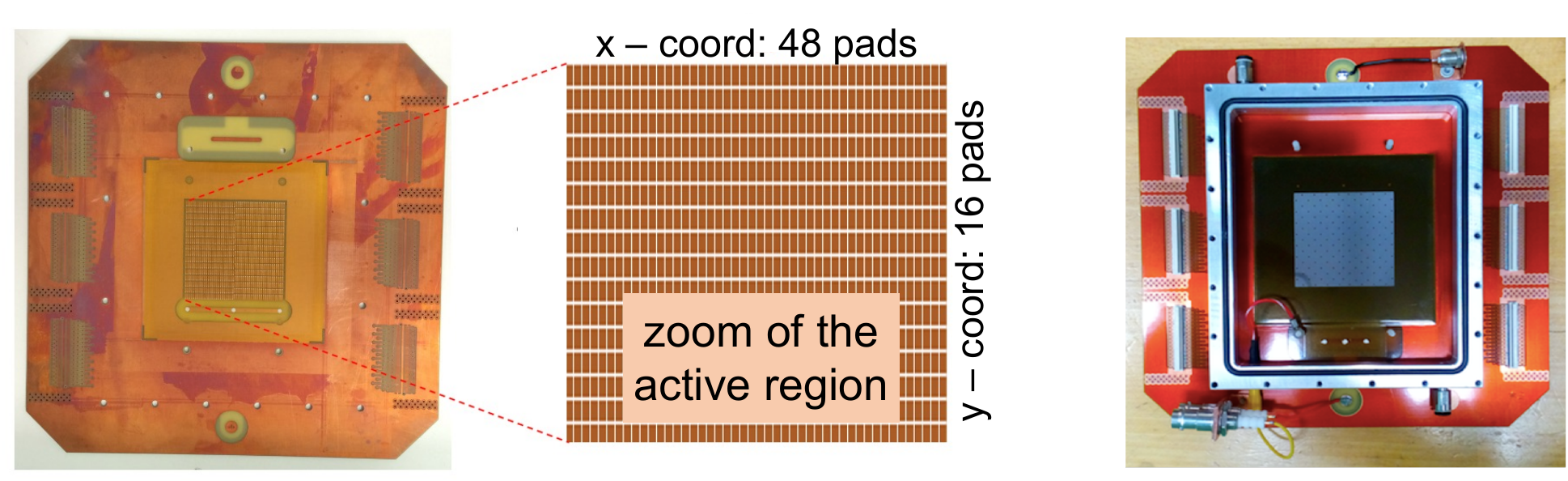}
\end{center}
\caption{Picture of the detector anode plane (left) with an expanded view of the pad structure (center).
Right: picture of the detector after the {\it{bulk}} process with the frame defining the gas enclosure; readout connectors are clearly visible at the border.}
\label{fig:fig1}
\end{figure}

The main idea that underlies our detector design is to reduce its occupancy in order to fulfill the requirement on the rate capability: this can be achieved designing 
a resistive Micromegas with small enough readout electrodes. At the same time, the resistive scheme has to guarantee a time stable operation (spark suppression) and fast charge evacuation at high particle rate. 
We have built and tested many high granularity Resistive Micromegas using different resistive schemes, in order to try different technologies and exploit all the advantages  of the different techniques that have been used.
All detectors presented in this paper share the same anode plane, segmented with a matrix of $48 \times 16$ readout pads. Each pad has a rectangular shape ($0.8 \times 2.8$ mm$^2$) with a pitch of 1 and 3 mm in the two coordinates. 
The active surface is $4.8 \times 4.8$ cm$^2$ with a total number of 768 channels and a density of about 33 readout elements per cm$^2$, routed off-detector for next signal processing stages.
The readout connectors are hosted on the border of the detectors where the front-end electronics is plugged-in.
Figure~\ref{fig:fig1} shows a picture of the anode plane. A resistive layer is put on top of this anode plane to act as a spark protection system. Different concepts of resistive layers have been implemented 
and will be discussed in this section. All MM detectors are assembled using the \emph{bulk} Micromegas process~\cite{bulk}, defining the 128 $\mu$m wide amplification gap with a metallic micro-mesh woven using stainless steel wires
having a diameter of 18~$\mu$m and opening windows of 45~$\mu$m. The mesh is supported by cylindrical insulating pillars  with a diameter of 300~$\mu$m  and an height of 128~$\mu$m (unless differently stated),
The detector is completed using a copper cathode to define the 5 mm wide conversion and drift gap.
\subsection{Resistive protection layer: different approaches}
The resistive layout is a crucial element of our Micromegas detectors. The detector performance has a strong dependence on its characteristics. 
We have implemented and studied several concepts of the spark protection resistive layers.  Two main schemes can be defined, 
even if some hybrid solution have also been exploited. In the following we describe all the resistive layouts that have been built and
tested, ordering them from oldest to newest one.
\begin{figure}
\begin{center}
\includegraphics[width=0.70\textwidth]{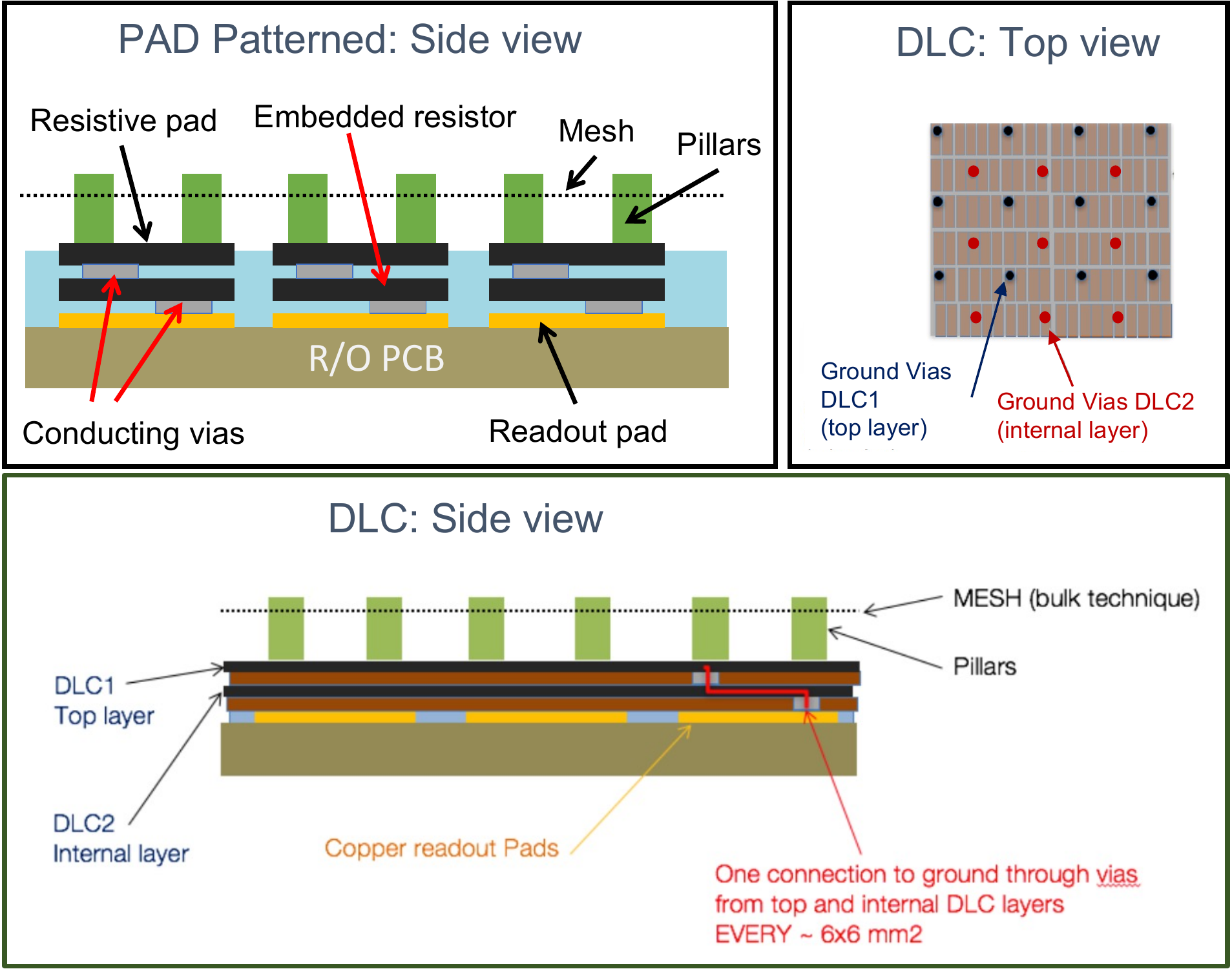}
\end{center}
\caption{Sketch of the pad-patterned embedded resistor layout (top left) and of the Diamond-Like Carbon layout (top right and bottom).}
\label{fig:fig2}
\end{figure}

\noindent {\bf{Pad-Patterned embedded resistors layout (PAD-P)}}

The first proposed detector~\cite{PaddyPaper} presents a pad-patterned (PAD-P) resistive layer, where each readout pad is overlaid by a resistive pad, both interconnected by intermediate resistors embedded in the structure.
It has been inspired by a similar development considered for the COMPASS experiment~\cite{Compass} (later abandoned) and for hadron sampling calorimetry~\cite{Chefdeville}. A schematic view of the layout is reported in Figure~\ref{fig:fig2} (left). It consists in stacking over each copper readout 
pad other two resistive pads (screen printed with a suitable resistive paste) interspersed with an insulating kapton layer; the outermost pad (facing the gas gap), the middle one and the innermost pad (the metallic one) 
are all electrically connected through staggered connection vias made by micro-holes filled with silver paste. 
With this solution the middle pad acts as a resistor, totally separated from the neighbors, whose value ranges between 5-7~M$\Omega$, depending on the paste resistivity; moreover
an almost uniform resistance to the anode pads is guaranteed, independently from the impact position of the electron avalanche on the innermost layer.

\noindent {\bf{Diamond-Like Carbon (DLC) layout}}

The second technique used to obtain the resistive coverage of the readout elements is referred in this paper as "DLC" and  takes its name from the use of the Diamond Like Carbon surface treatment, that consists in sputtering carbon (evaporated from a graphite target) on a kapton foil, obtaining a uniform resistive layer. The layout schema, that has been inspired by the technique used for the $\mu$-RWELL detector~\cite{uRwell}, is  shown in fig.~\ref{fig:fig2} (top right and bottom): two DLC foil, covering the readout plane, are interconnected through staggered conductive vias, providing the charge evacuation through their resistive surface, such offering a mean impedance that is almost independent from the charge collection position. 
The pitch of the conductive vias network represents a relevant parameter of the detector, as well as the surface resistivity of the DLC foils. 
For this reason two prototypes have been built and tested with this technique using different resistivity: the first one with average resistivity 
of about 50~M$\Omega/\square$, and the other with foils with about 20~M$\Omega/\square$ referred to as DLC50 and DLC20, respectively.
Moreover, for both detectors, the active plane has been divided in two halves for testing purposes, with a different pitch of the conducting vias through the DLC layers: 6 mm and 12 mm respectively.
In order to distinguish these two regions, the suffixes  "\textit{\minus 6mm}"  and  "\textit{\minus 12mm}" are added to the corresponding name.
\newpage
\noindent {\bf{DLC with the sequential build-up technique (SBU)}}

\begin{figure}[htbp]
\begin{center}
\includegraphics[width=0.45\textwidth]{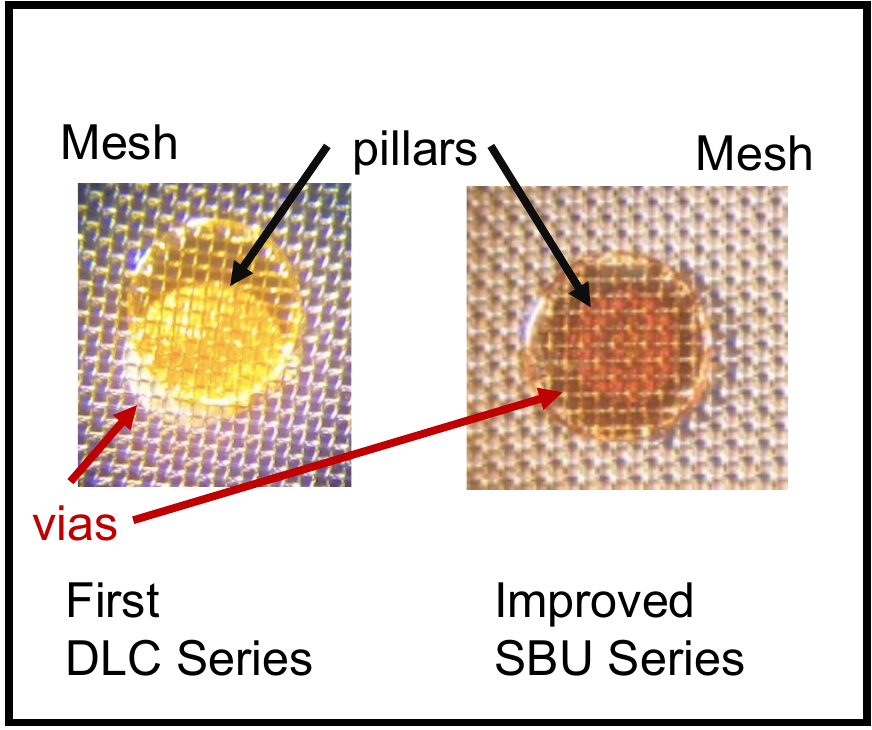}
\end{center}
\caption{Left: example of pillar (uppermost yellow circle) in DLC20/50 prototypes, not well centered with the conductive vias. 
The mesh layer can also be seen. Right: example of good alignment between pillar and conduction vias in SBU prototypes.}
\label{fig:fig3}
\end{figure}

In order to further improve the precision of the construction process, a new technological solution for the production of the resistive layers has been introduced making use of copper clad DLC foils. 
This technique has been named Sequential Build Up (SBU)~\cite{iodiceSBU} and it uses photolithography  to precisely locate the conductive vias and align them below the pillars
as shown in Figure~\ref{fig:fig3} right in comparison with a misaligned via-pillar pair observed in one of the first standard DLC prototypes (Figure~\ref{fig:fig3} left).
Detectors produced with SBU technique are quite similar to standard double DLC detectors but they have a better stability by avoiding a conducting surface exposed to the amplification region which can prevent discharges.
Moreover this solution is much more convenient than the simple double DLC process described before because this building technique is fully compatible with standard PCB processes, 
significantly facilitating the technological transfer of the production. 
Three prototypes have been built with the SBU technique, referred in the following sections as SBU1, SBU2 and SBU3. All of them, have been built using a configuration with the 6~mm pitch grounding vias, adopted in the full area.
The first two detectors were built with a $\sim$5~M$\Omega/\square$ resistive layer facing the amplification region and an innermost layer with a resistivity of $\sim$35~M$\Omega/\square$.
The third prototype, SBU3, has both DLC foils with a mean surface resistivity of about 30~M$\Omega/\square$ and the copper readout pads located between the two resistive DLC foils to improve the capacitive coupling in signal induction.

\noindent {\bf{Hybrid layout (PAD-H)}}

\begin{figure}[htbp]
\begin{center}
\includegraphics[width=0.80\hsize]{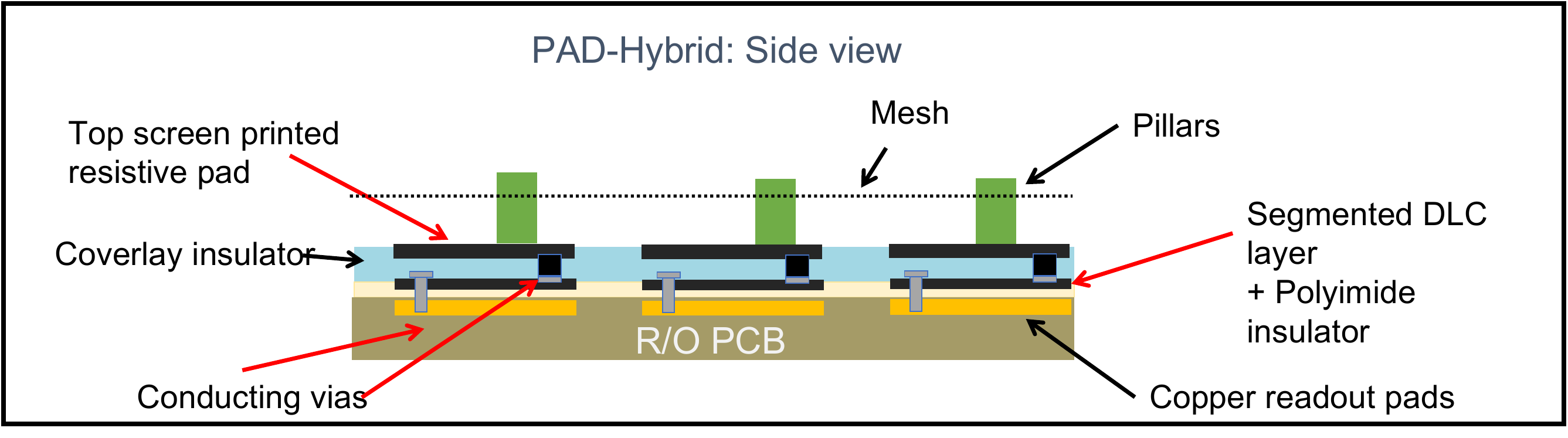}
\end{center}
\caption{Schematic view of the PAD-H prototype.}
\label{fig:fig4}
\end{figure}

This fourth kind of detector can be considered a technological improvement of the Pad-Patterned embedded resistors layout.
This configuration uses a uniform DLC layer for the inner layer and screen-printed resistive pads for the outer layer. 
Different from the DLC and the SBU schemes, in the PAD-H configuration the carbon layer of the DLC foil is patterned in pads by means of a proper etching procedure.
A schematic cross-section of the detector is shown in Figure~\ref{fig:fig4} top, with  indication of the components of the stack.  
The size of the DLC pads can be equal to the one of the screen-printed pads to maximize the number of charge evacuation paths, 
or to a multiple of them to simplify the construction at the cost of longer evacuation paths. 
In our prototype we opted for the first solution, leading to the same connection scheme, between the inner and the outer layers, of the PAD-P detectors.

\noindent {\bf{DLC-Strip layout}}

\begin{figure}[htbp]
\begin{center}
\includegraphics[width=0.80\hsize]{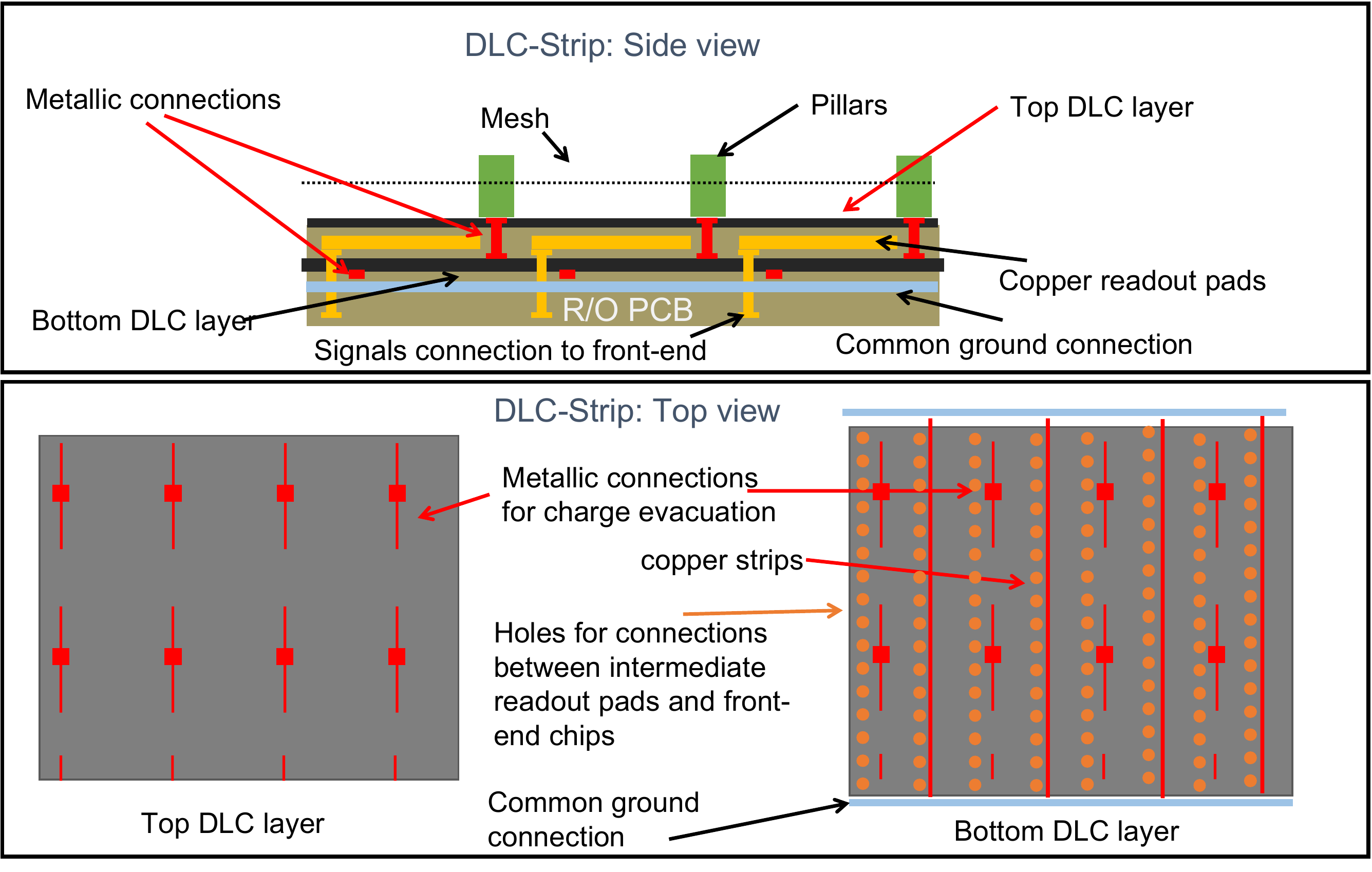}
\end{center}
\caption{Schematic view DLC-Strip detector. The bottom plot shows the layout of the metallic connection on both resistive layers and the necessity to cover them with elongated pillars.}
\label{fig:fig5}
\end{figure}

The latest prototype that has been built is really quite different from the previous ones because of many changes in the design of both resistive protection and signal formation.
The first main feature of this detector is that the readout pads are located in-between the two uniform DLC resistive layers in order to improve the capacitive coupling for signal induction,
as shown in Figure~\ref{fig:fig5}. Another important change with respect to the former prototypes is that metal connection strips are used to connect each other the DLC layers, for this reason we will refer to this design scheme as  DLC-Strip layout.
The reason for this solution lies in the fact that metallic connections provide a much more reliable solution for charge evacuation than conductive vias obtained with a silver-loaded polymer. 
Moreover, as illustrated in Figure~\ref{fig:fig5} bottom, the metallic connection strips on both resistive layer
divide the detector surface into different areas where the charge is collected more quickly and more efficiently, thus making the evacuation independent of the width of the irradiated surface.
Also in this case, to avoid the development of intense discharges top surface of these metallic connections must be completely insulated by the gas gap, and that is obtained as usual by covering 
the exposed conductive material with pillars. Anyway in this case they need to be elongated to a length of 5~mm, as shown in Figure~\ref{fig:fig5} bottom left, thus reducing the geometrical acceptance of the detector.

\subsection{Detector characterization and performance studies with different gas mixtures}

All the detectors have been characterized and extensively tested  in order to measure their performance in terms of charge up behavior, gain, rate capability, 
energy resolution,  efficiency and spatial resolution and detector stability.  Measurements using radioactive sources ($^{55}$Fe), and 8 keV photons from a Cu X-rays gun have been performed at the 
GDD (Gas Detector Development) laboratory of the RD51 \cite{RD51} Collaboration at CERN. 
Spatial resolution has been measured in dedicated test beam activities at CERN beam lines, using high energy muon and pions, looking at the residuals distribution with respect an external tracking system. 
More irradiation tests have been performed at PSI using intense pion beams and at CERN in the GIF++ facility \cite{GIF}.
Two kind of gas mixtures have been tested: the binary Ar:CO$_2$ gas mixture (93:7) as a standard reference, and a ternary mixture Ar:CO$_2$:iC$_4$H$_{10}$ (93:5:2), at a gas flow of few renewals per hour, for a typical gas volume of 0.2~l.
A nominal electric field in the drift region of 60 V/mm has been used as a reference for all measurements, unless differently stated, thus optimizing the mesh electron transparency.

In this section a summary of the most relevant results is reported: a complete and detailed review of them can be found in \cite{PaddyPaper, iodiceSBU, Elba2018, ichep2020}.

\subsection{Gain, Charging up effect and rate capability}

\begin{figure}[htbp]
\begin{center}
\includegraphics[width=0.80\hsize]{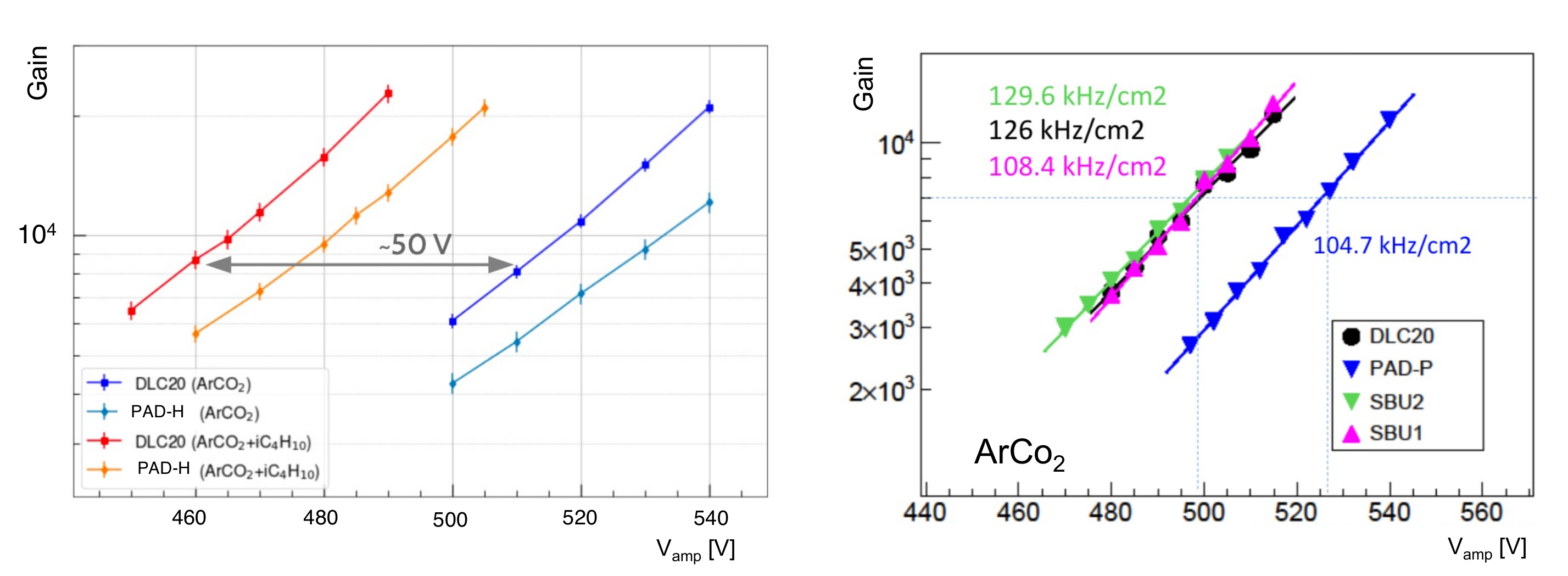}
\end{center}
\caption{Gain vs amplification voltage for PAD-H and DLC detectors in Ar:CO$_2$ and Ar:CO$_2$:iC$_4$H$_{10}$  (left) and for PAD-P, DLC and SBU detectors in Ar:CO$_2$ (right).}
\label{fig:fig6}
\end{figure}

Detector gain has been measured using different methods, depending on the range of the rate of incoming radiation seen by the detector.
Al low rate (below few hundreds kHz, when detector is exposed to $^{55}$Fe source or to 8 keV X-Rays)  the detector current is measured from readout 
pads with a pico-ammeter and signal rate from the mesh is counted directly or through a MultiChannel Analyzer (MCA).
Figure~\ref{fig:fig6} right shows the gas gain as function of the amplification voltage of a PAD-P, a DLC and two SBU detectors. 
DLC and SBU, which both have an external uniform DLC resistive layer, show the same gain, thus demonstrating the high level of uniformity reached in the production process.  
The PAD-P detector shows a lower gain, by about a factor 2, only partially justified by the larger charge-up. 
The main difference is attributed to the different field configuration as a consequence of the different layouts of the external resistive layers.
The same difference was observed comparing the DLC20 detector with the PAD-H, as shown in  Figure~\ref{fig:fig6} left.
In the same plot are also reported gain curves of the two detectors measured with the Ar:CO$_2$:iC$_4$H$_{10}$ mixture in the fraction 93:5:2. 
The introduction of 2\% of isobutane lead to a gain increase of a factor about 4 with respect to the Ar:CO$_2$ mixture, owing to the higher Penning transfer. 
It has been observed that adding isobutane drastically reduces the intensity of discharges, thus ensuring a much more stable operation working point. 
The mixture with the addition of 2\% of isobutane allows then to operate the detector at lower voltage to reach the same gain. 

\begin{figure}[htbp]
\begin{center}
\includegraphics[width=\linewidth]{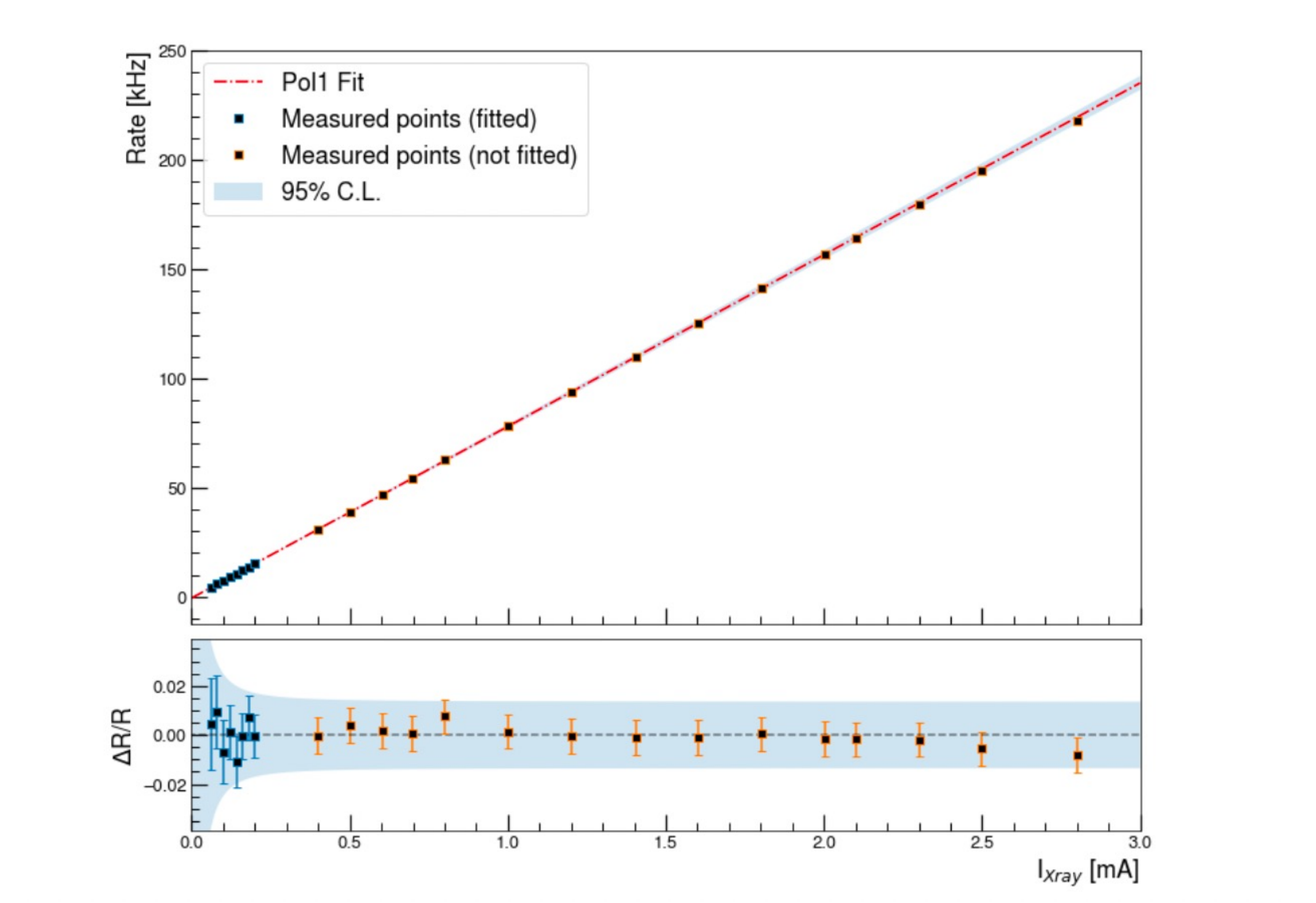}
\end{center}
\caption{PAD-H detector response as function of the current of the X-ray gun: the detector shows an excellent linear behavior up to 200 kHz.}
\label{fig:fig7}
\end{figure}

At higher rate values, which can be reached using the X-rays gun exploiting the full range of the current flowing in its filament, a direct measurement of the detector rate is
no longer reliable due to events pile-up: then, for this reason, detector rate has been estimated using an extrapolation at higher current values in the linear relation 
between detector response and incoming flux of particles. 
This relation has been verified over two orders of magnitude and it is shown in Figure~\ref{fig:fig7}: measured points have a relative uncertainty less than 2\% with respect the extrapolation taken at very low value of incoming radiation 
and they are within the 95\% CL prediction over the exploited range up to 200 kHz. At even higher values, up to several tens of MHz, any other deviation from this linear behavior is then attributed to a detector gain loss. 

The variation in time of the gain in MPGD when exposed to intense radiation fluxes is a well known effect and has been 
observed by many authors  \cite{chargeup1,chargeup2}. This phenomenon is due to to the charging up of the dielectric material in the detector structure and it can produce 
either an increase or a  reduction of the gain, depending on the field configuration: both effects can be present in the same structure although with different time scales. 
We have measured a different behavior of the detectors with the upper resistive layer segmented in pads (like the PAD-P series)  with respect to detectors with an uniform DLC layer. (as the DLC and SBU series).
\begin{figure}[htbp]
\begin{center}
\includegraphics[width=\linewidth]{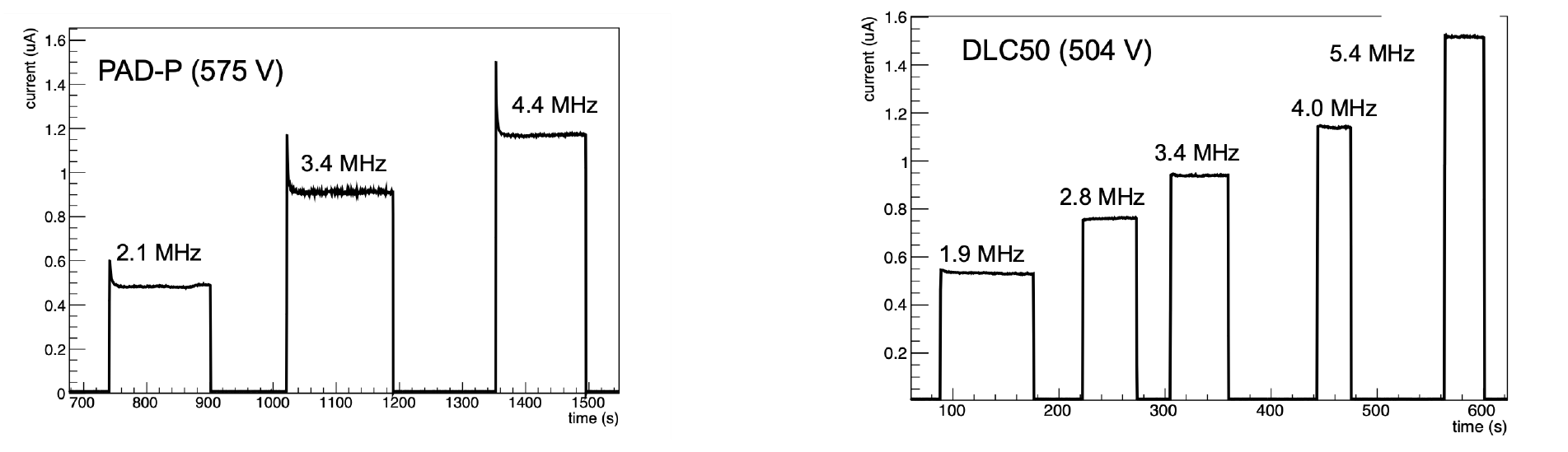}
\end{center}
\caption{Charge up for the PAD-P (left) and the DLC (right) detectors irradiated with X-rays.}
\label{fig:fig9}
\end{figure}
As shown in Figure~\ref{fig:fig9}, the current as function of the time has been recorded for a PAD-P and a DLC detectors, when irradiated with X-rays, changing the irradiation rate in a range of a few MHz/cm$^2$ by discrete steps.
The PAD-P detector shows a fast gain reduction of the order of 15-20\% within a time scale of a few seconds, and we can interpret this result as an effect due to the charging-up of the exposed dielectric surface.
This effect is almost negligible at the same short time scale for detectors presenting an uniform resistive DLC layer facing the gas amplification region (DLC and SBU series), with a gain reduction of less than few percent,
PAD-H detectors showed a charging up effect similar to the PAD-P series.
The opposite effect was observed during long-term (about 10~h) irradiation with a high intensity pion beam at PSI.  Results are shown in Figure~\ref{fig:fig10} left: after a short initial period when the PAD-P detector suffers a 
decrease of the current, different from the others, all the detectors show an increase of the current in the long run. 
Figure~\ref{fig:fig10} right illustrates the peculiar behavior of the DLC-Strip prototype: it presents no detectable gain drop at short time scale, as all the prototypes with a top uniform resistive layer (DLC and SBU series) but
it shows a gain increase of $\sim 3\%$ after a few minutes even at an hit rate of about a few hundreds of  kHz/cm$^2$ under X-ray irradiation.

\begin{figure}[htbp]
\begin{center}
\includegraphics[width=\linewidth]{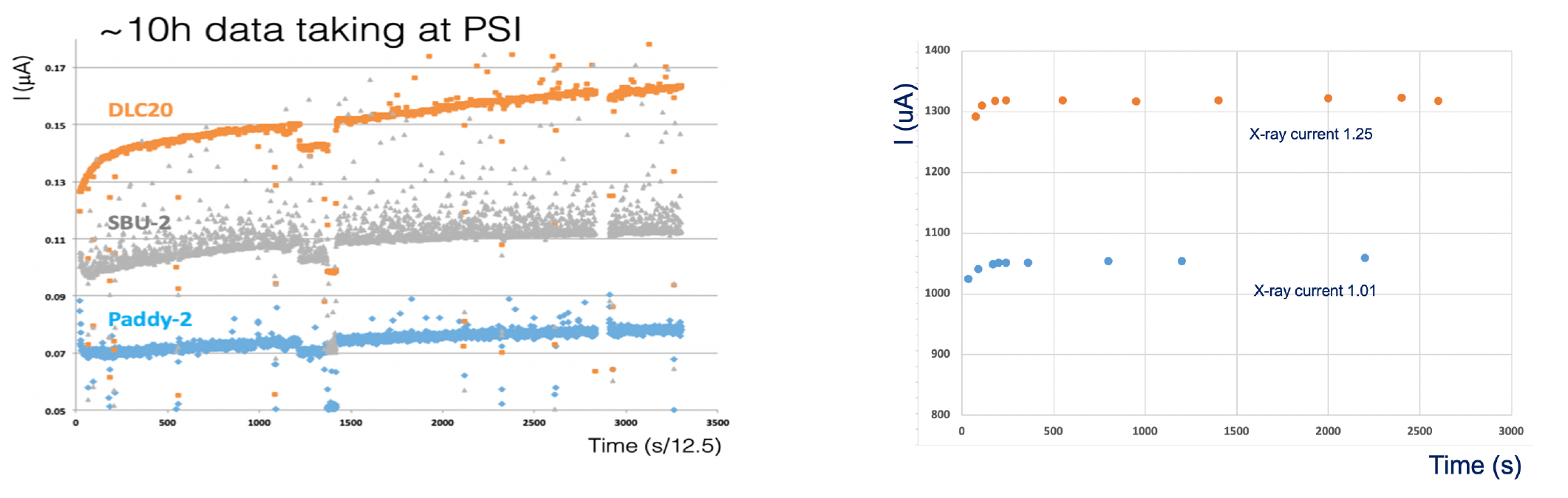}
\end{center}
\caption{Charge up for the PAD-P, DLC and SBU at PSI pion beam (left) and of the DLC-Strip (right) irradiated with X-rays.}
\label{fig:fig10}
\end{figure}

\begin{figure}[htbp]
\begin{center}
\includegraphics[width=0.95\textwidth]{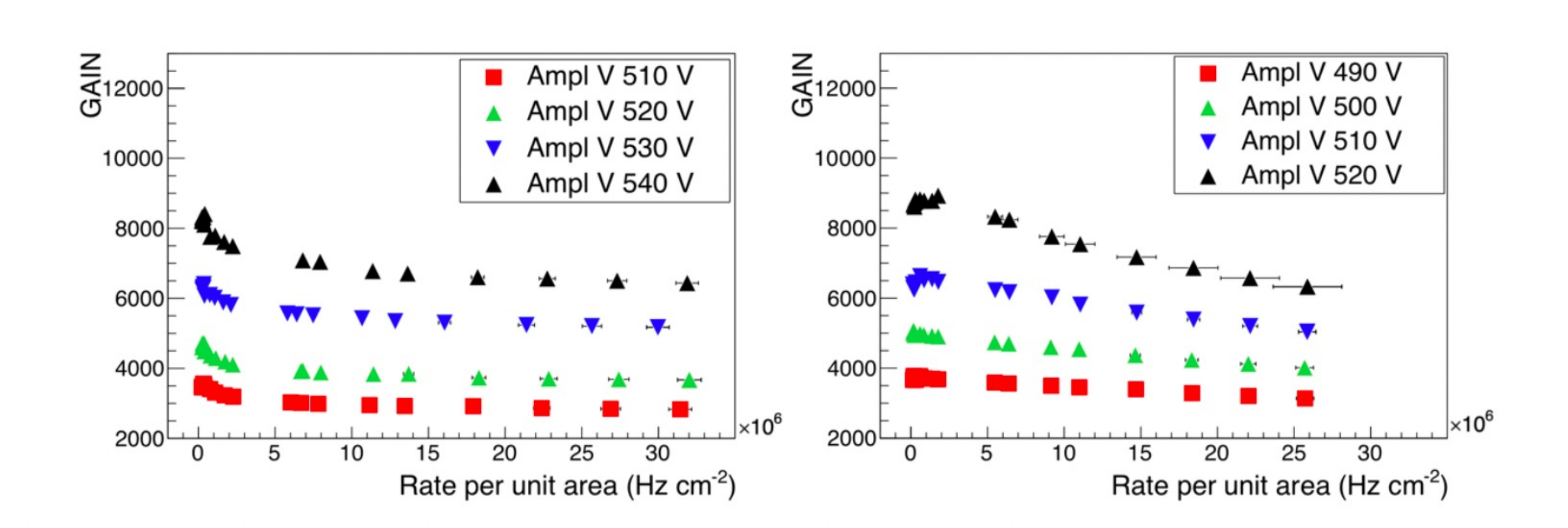}
\end{center}
\caption{Gain of the PAD-P (left) and DLC20-6mm (right) in the range of rates up to 30 MHz/cm$^2$ for different values of the amplification voltage, measured with X-rays.}
\label{fig:fig11}
\end{figure}

To assess the rate capability of the detectors, their gain has been measured over a wide range using Cu target X-rays.
Figure~\ref{fig:fig11} reports the gain of the PAD-P and DLC-20 detectors in the range from a few kHz/cm$^2$ up to 30 MHz/cm$^2$ for different values of the amplification voltage.
As already stated in the previous section, the pad patterned prototype shows a significant gain drop at lower rates (below 5 MHz/cm$^2$ ) dominated by charging up effect, up to about 20\% at 10 MHz/cm$^2$ at 530~V,
while it has a negligible ohmic voltage drop in range of rates from 10 to 30 MHz/cm$^2$. 
On the contrary, the DLC-20  detector shows an almost constant gain in the range below few MHz/cm$^2$ and than, at higher values, a significant ohmic voltage drop, with a relative drop
of  about 20\% at 20 MHz/cm$^2$ at 510~V. This behavior is common to all the DLC series detectors, including the SBU type; moreover the detectors with uniform resistive layer have 
systematically a gain higher than PAD-P (at low/moderate rates) for the same value of the amplification voltage: 
this is due to a more uniform electric field in the amplification gap with respect to PAD-P. For this reason, we usually operated the two kind of detectors at a sligtly different voltage values, in 
order to keep them at the same gain values when comparing their performances.  

\begin{figure}[htbp]
\begin{center}
\includegraphics[width=1.0\hsize]{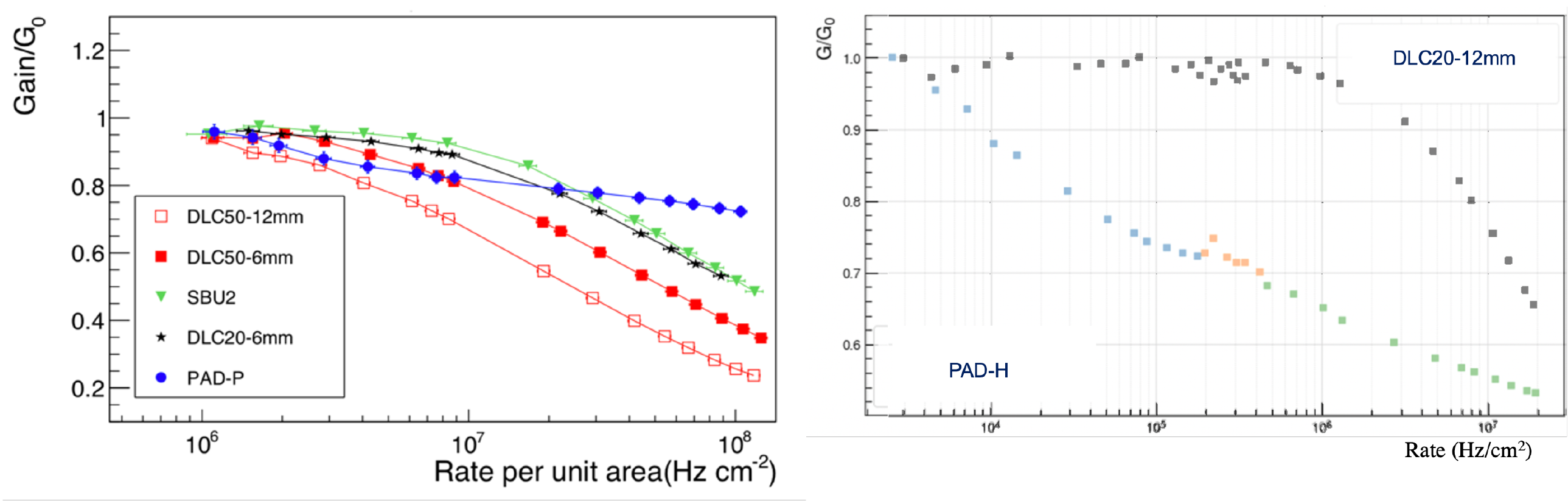}
\end{center}
\caption{Left: Dependence of the gain of the PAD-P, DLC and SBU detectors, normalized to their value at low rates, as a function of the X-Rays hit rates. The amplification voltage was set to have a gain about 6500 at 100~kHz/cm$^2$ for all the detectors. Right: direct comparison between the PAD-H and DLC20 detectors operated at approximately the same initial gain.}
\label{fig:fig12}
\end{figure}

Figure~\ref{fig:fig12} left reports a comparison between the different types of detectors: relative gain loss of PAD-P, DLC and SBU prototypes has been measured as a function of the hit rates in the range 1-100 MHz/cm$^2$. 
As previously mentioned, the different behavior of the PAD-P prototype is evident: its gain drop is dominated by the charge up, increasing with the rates and almost saturating at 20 MHz/cm$^2$, where the gain drops by about 20\%.
With a current larger than 0.5~$\mu$A per pad, the voltage ohmic drop is contributing only for very high rates,  up to a total gain drop of about 30\% at 100~MHz/cm$^2$. 
Instead, all the detectors belonging to DLC or SBU series have the same behavior  at rates above few MHz/cm$^2$ , but it dramatically depends on the value of the surface resistivity:
the DLC50 prototype is more severely affected by the ohmic voltage drop and the gain is significantly reduced, as expected because of its higher resistivity. Moreover the configuration with 6~mm pitch grounding vias has a 
more reduced gain loss with respect to the 12~mm configuration, because there the charge sees a much higher impedance to ground.
The DLC20-6mm and SBU2 detectors, which share the same configuration and have a quite similar value of the DLC surface resistivity show a similar 
behavior at high rates, with a gain drop similar to PAD-P at about 20~MHz/cm$^2$, further reduced up to about 50\% at 100~MHz/cm$^2$. 
 
In the same picture, but on the right side, there is the gain drop of PAD-H and  DLC20 detectors, 
reported as a function of the hit rates and normalized to their value at low rates, in the range from few kHz/cm$^2$ up to few tens of MHz/cm$^2$. 
Data have been taken irradiating detectors with X-Rays almost uniformly in a circular area of 0.79 cm$^2$ and operating them at approximately the same gain. 
Even in this case, the two detectors show a quite different response to the increase of the rate. PAD-H undergoes a rapid loss of gain even at few kHz, with an apparent exponential drop of  $\sim$25\% within two orders of magnitude. 
The behavior of this detector confirms what has been measured for other pad patterned detectors and seems to be largely dominated by the charging-up.  Anyway it can operate with still significant gain even at very high rates. 
By contrast, the prototype with the DLC uniform layer resistive scheme shows constant performances for an extended range of rates. It shows variations of the gain smaller than 10\% up to few MHz/cm$^2$ of incident radiation.

\begin{figure}[htbp]
\begin{center}
\includegraphics[width=0.75\linewidth]{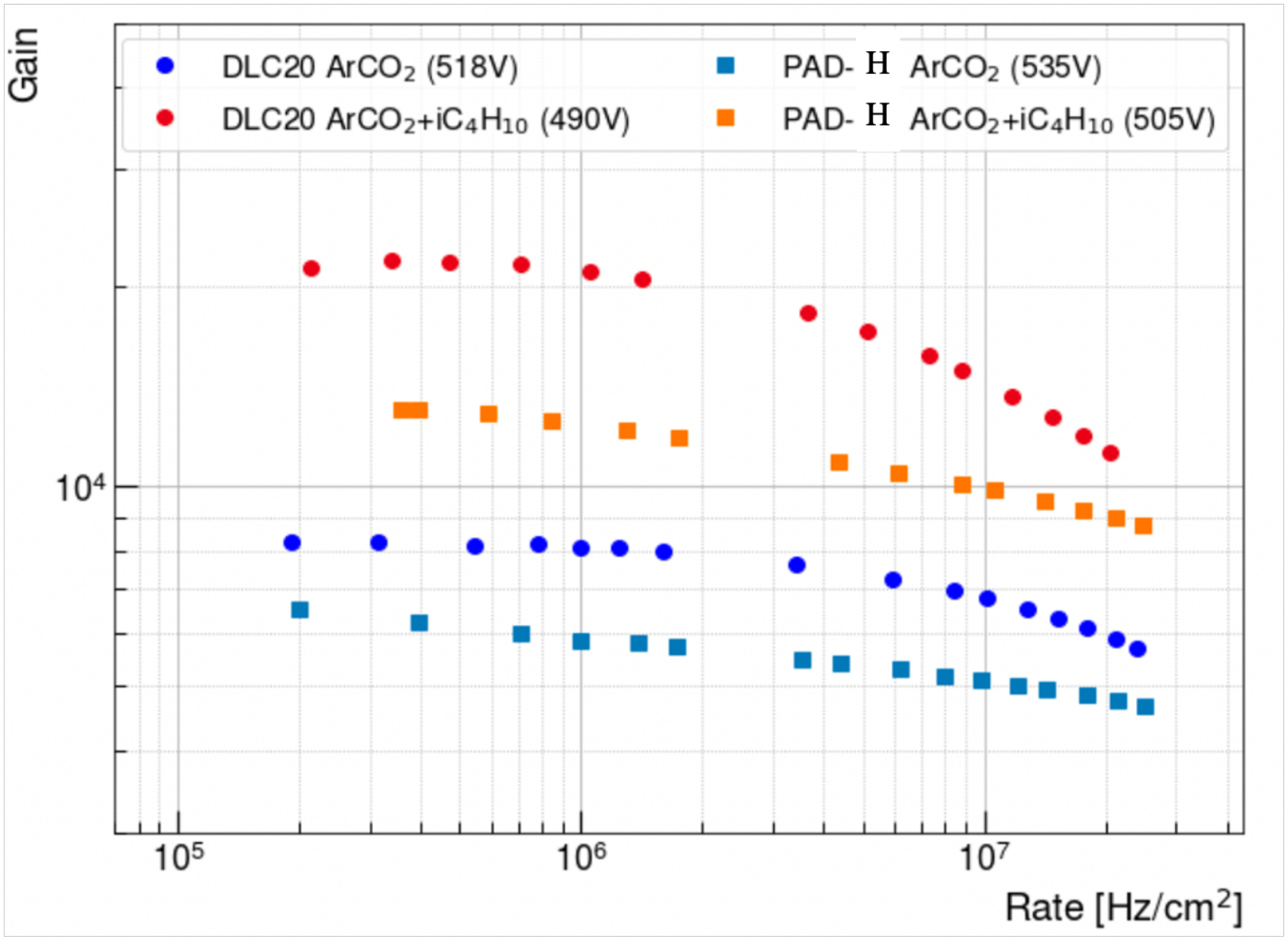}
\end{center}
\caption{Gain as function of the hit rate for the DLC20 and PAD-H detectors operated with Ar:CO$_2$ (93:7) and Ar:CO$_2$:iC$_4$H$_{10}$ (93:5:2) gas mixtures.}
\label{fig:fig13}
\end{figure}

The performances in rate capability can be enhanced operating the detector with isobutane enriched Ar:CO$_2$:iC$_4$H$_{10}$ (93:5:2) gas mixture. As previously noted, (see Figure~\ref{fig:fig6}) this ternary mixture leads 
to a gain increase of a factor about 4 with respect the binary Ar:CO$_2$ (93:7) gas mixture thus allowing to stable operate the detectors with gain above 10$^4$ up to extremely high rates of the order of 10MHz/cm$^2$ and more.
Results are reported in Figure~\ref{fig:fig13}.

\subsection{Energy resolution}
Several test beam campaigns with particle beams at the CERN SPS H4 line and at PSI have been conducted to measure 
the spatial resolution and efficiency of the detectors. During fall 2021 measurements have been performed also at CERN GIF++ facility~\cite{GIF} where a muon beam is 
available together with 662~keV photon background from a $^{137}$Cs source of 16.6~TBq activity.
Data have been acquired with APV-25 hybrid chips and read out with the SRS system~\cite{APVSRS}. 
A scintillator hodoscope has been used for triggering incoming particles and two double view bulk resistive Micromegas detectors with strip segmented anode have been used as tracking chambers.
All the presented results have been obtained with muon or pion beams perpendicular to the detector surface.
Clusters have been reconstructed combining the charge information from neighboring fired pads and their position is obtained 
as the charge weighted centroid of the fired pads that pass minimal quality cuts, used to clean up the signals from front-end electronic noise.

\begin{figure}
\begin{center}
\includegraphics[width=0.95\textwidth]{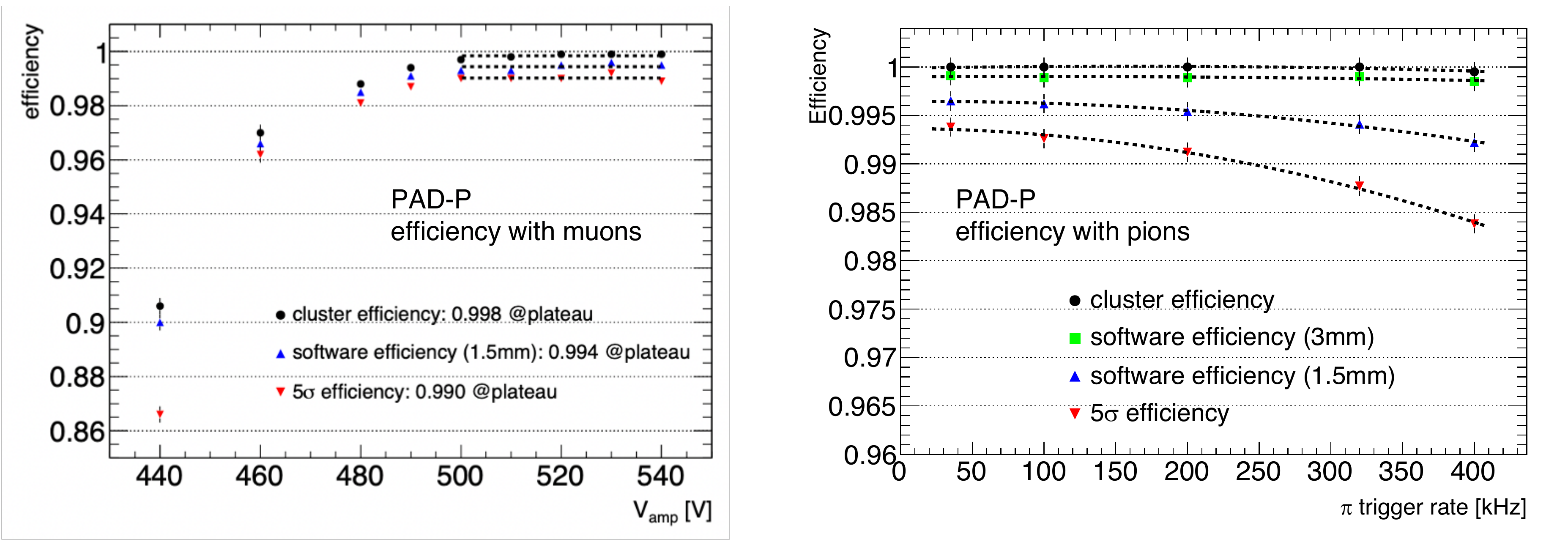}
\end{center}
\caption{Left: PAD-P efficiency vs Amplification voltage, as measured with a high energy muon beam. Right: PAD-P efficiency vs pion beam rate.}
\label{fig:fig14}
\end{figure}

Three types of efficiency have been defined: a) cluster efficiency: an event is classified as efficient if a good reconstructed cluster is found; b) software efficiency: an event is efficient if a good reconstructed cluster is 
found within an acceptance window of 1.5 mm around the position of the extrapolated track; c) 5$\sigma$ efficiency: an event is efficient if a good reconstructed cluster is found within five times 
the width of the residuals distribution around the extrapolated track.
In Figure~\ref{fig:fig14} left the efficiency of the PAD-P detector as function of the amplification voltage is reported.; on the right side of the same picture, the plateau efficiency of the PAD-P detector is measured as a function 
of the pion beam rate. Results confirm that even with the most stringent requirement, the detector efficiency  is well above 98\%. It is worth to note that for this chamber the pillars cover about 2.8 \% of the active area surface.

\begin{figure}
\begin{center}
\includegraphics[width=0.80\hsize]{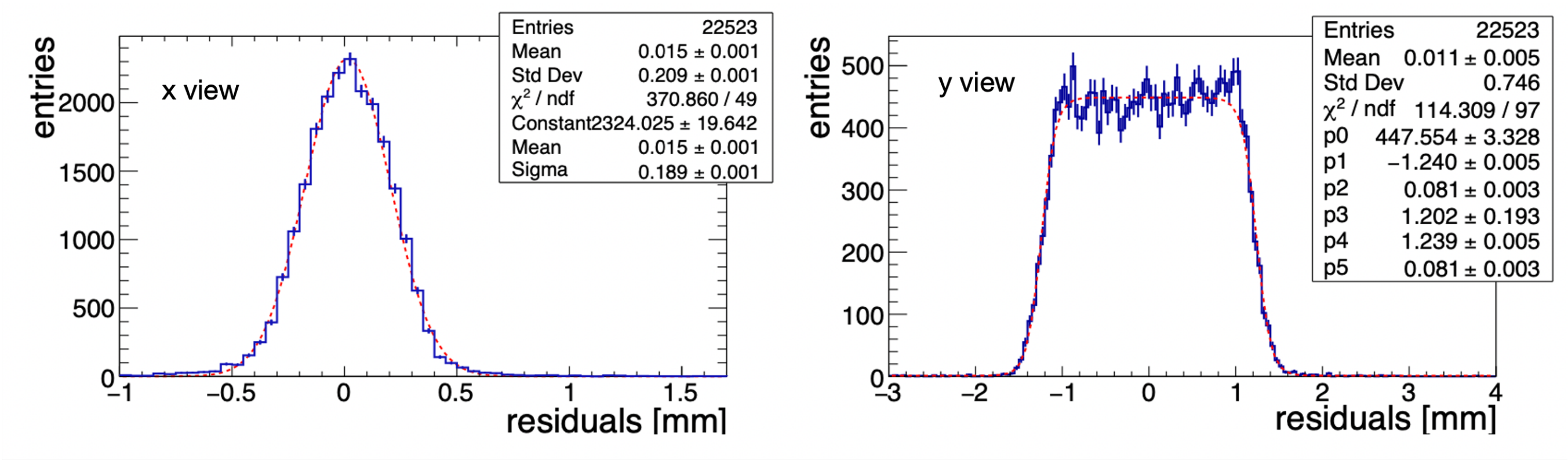}
\end{center}
\caption{Distribution of the residuals between the reconstructed cluster position on the PAD-P detector and the extrapolated track position measured with an external tracker, for the x- (left) and y-coordinate (right).}
\label{fig:fig15}
\end{figure}

The spatial resolution has been evaluated comparing the cluster position reconstructed on the detector and the expected hit position obtained from tracking detectors; results for PAD-P
detector are reported in Figure~\ref{fig:fig15}.  The spatial resolution measured with CERN pion beam in the precise coordinate (x, with pad size of 0.8~mm and a pitch of 1~mm) is 190~$\mu$m. 
In the second coordinate (y with a pad readout pitch of 3~mm) the residuals appear quite uniformly distributed with a FWHM of about 2.4~mm, 
as expected from the large pad size. 

\begin{figure}
\begin{center}
\includegraphics[width=0.80\hsize]{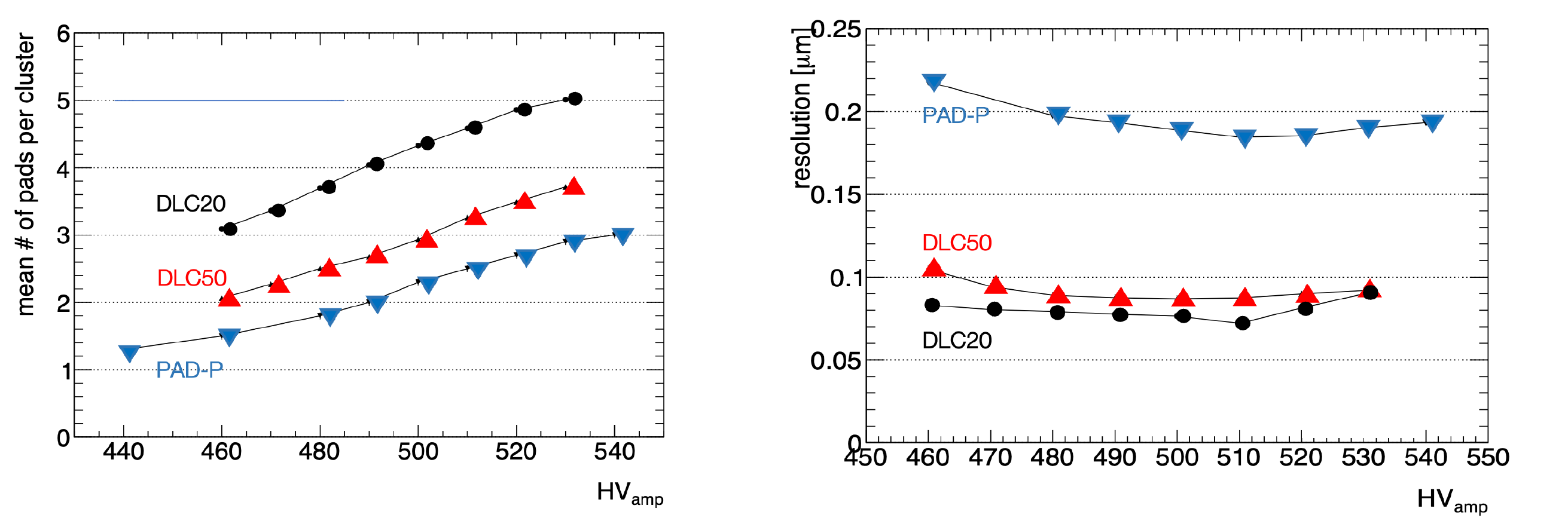}
\end{center}
\caption{Cluster dimension (left) and spatial resolution (right) for the PAD-P, DLC20 and DLC50 detectors measured with pion beam at the CERN SPS.}
\label{fig:fig16}
\end{figure}

Of course the main parameter affecting the spatial resolution is the readout pad dimension, that is the same all the detectors; anyway the configuration of the resistive layer plays a significant role because it affects the spread of 
the induced charge over the surface. 
In detector with uniform layers (DLC, SBU) the induced charge spreads over more pads, leading to larger average dimension 
of the reconstructed cluster and a more precise centroid reconstruction. A lower resistivity of the external carbon layer goes in the same direction.
This effect is clearly visible in Figure~\ref{fig:fig16} where the cluster size (left) and the spatial resolution (right) are reported as function of the amplification voltage for the PAD-P, DLC20 and DLC50 detectors.
DLC detectors, with a uniform resistive layer has a larger cluster size and, correspondingly, a spatial resolution better than 100~$\mu$m in the x-coordinate, while the 
PAD-P detector, with segmented resistive pad in the external layer, has a smaller cluster dimensions and a slightly worst spatial resolution. Moreover 
DLC20 detector shows a better spatial resolution with respect to DLC50 because of its smaller surface resistivity, as expected.

\begin{figure}
\begin{center}
\includegraphics[width=0.90\textwidth]{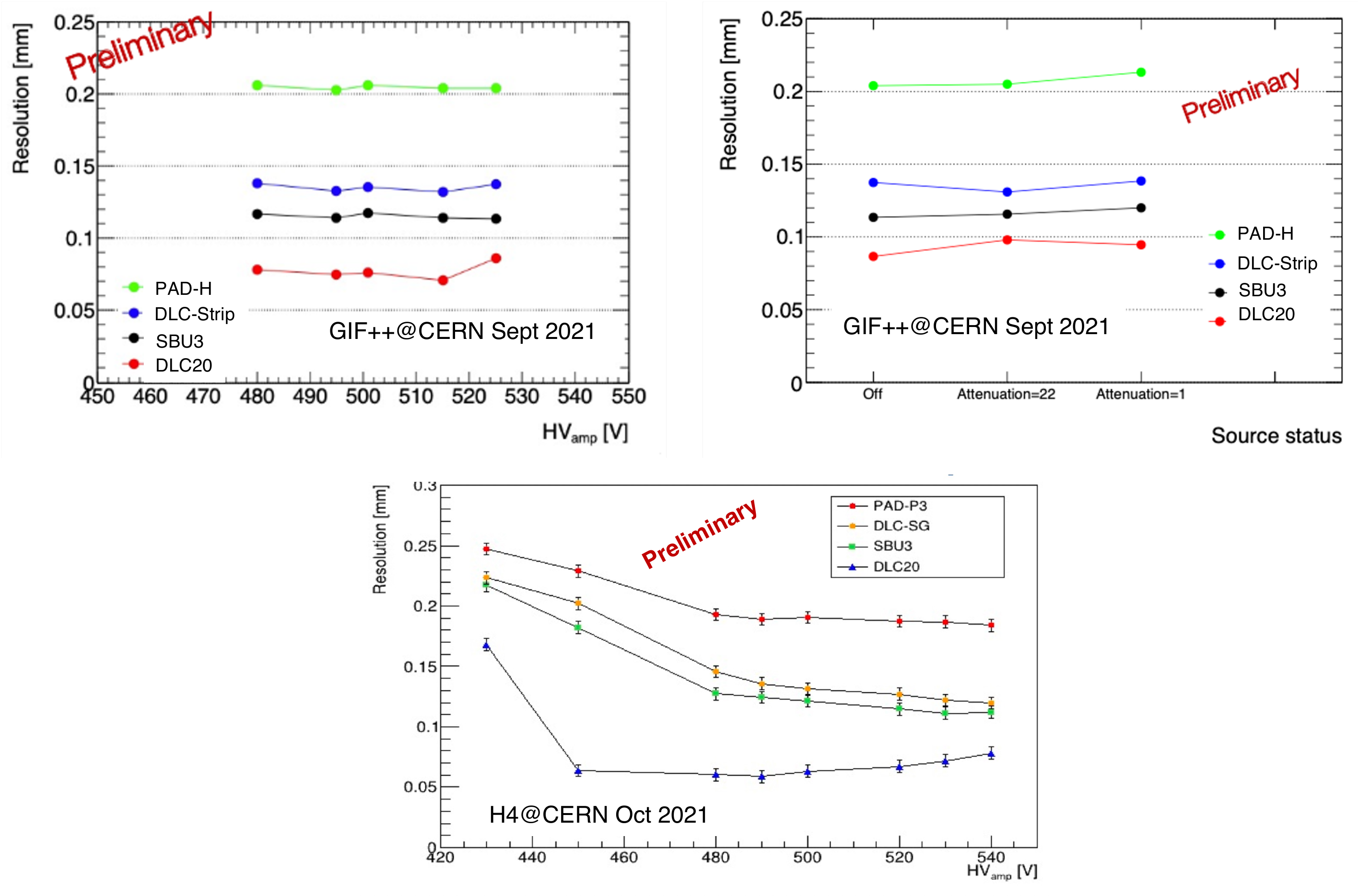}
\end{center}
\caption{Top: Spatial resolution measured at the GIF++ facility at CERN as function of the amplification voltage (left) and of the $^{137}$Cs source status (right), for the PAD-H, DLC20, DLC-Strip and SBU detectors.
Bottom: Spatial resolution measured at H4 at CERN with muon beam as function of the amplification voltage.}
\label{fig:fig17}
\end{figure}

Preliminary results of the fall 2021 test beam campaign are reported in Figure~\ref{fig:fig17}:  spatial resolution of PAD-H, the DLC-Strip, SBU3 and the DLC20 detectors, measured with the same technique mentioned above, is shown.
On the top left part of the figure the spatial resolution as function of the amplification voltage is represented: as expected the DLC detector, with uniform 
layer of lowest resistivity (DLC20), has a resolution of about 80~$\mu$m, in agreement with older measurements, 
while the pad-patterned device (PAD-H) shows a resolution of about 200~$\mu$m, that is compatible with PAD-P resolution.
The spatial resolution was also measured with increasing photon background. 
Results are shown in Figure~\ref{fig:fig17} top right where the x-axis reports the status of the GIF++ gamma source, ranging from "Off" , that means no photon flux 
was present, to "Attenuation=1" that means full source, corresponding to approximately few tens of kHz/cm$^2$. An intermediate value in this scale is reported with results obtained 
with a set of absorbers in front of the GIF++ source providing a total photon flux reduced by about a factor 22 with respect to full source. 
No degradation of the tracking performance have been observed for any of the tested detectors for the full range of the GIF++ photon flux.  
This result confirms, once more, the suitability of these detector as tracking devices in conditions with high background.
The bottom plot  in Figure~\ref{fig:fig17} shows the same detectors tested 
few weeks later using muon beam in H4 experimental area @ CERN. Results are fully compatible with the previous ones, thus demonstrating the stability of detector performances
in different experimental conditions.

\subsection{Detector stability}

\begin{figure}
\begin{center}
\includegraphics[width=\textwidth]{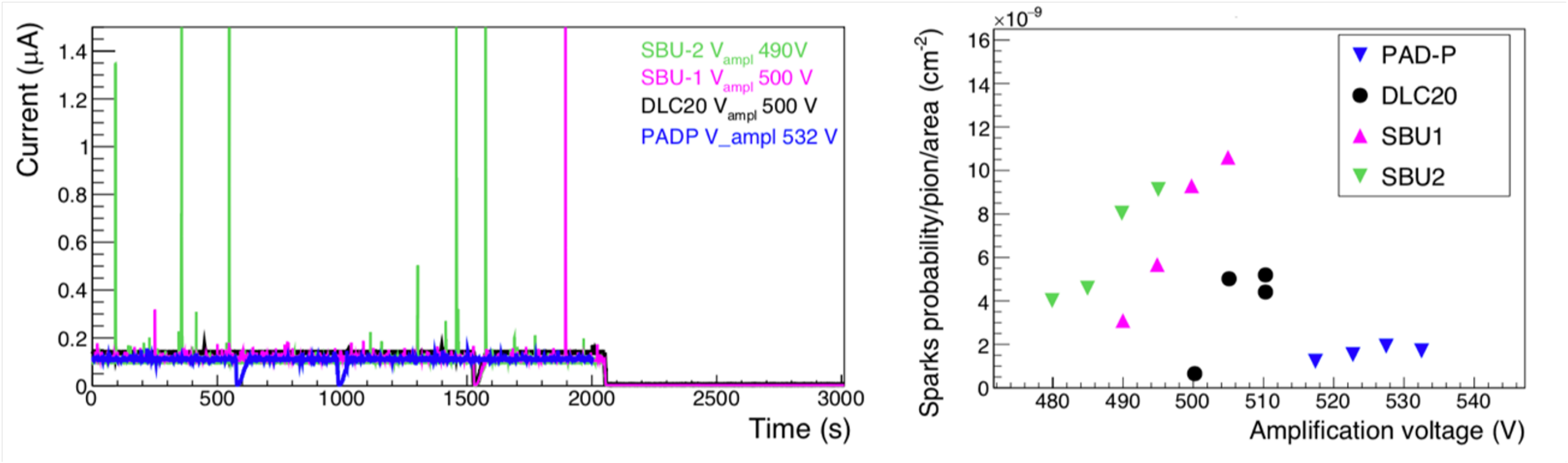}
\end{center}
\caption{Detector stability with a pion beam of 350 MeV/c. Left: current as a function of time under a particle rate of about 100~kHz/cm$^2$. Right: spark probability density per pion.}
\label{fig:fig18}
\end{figure}

The spark probability of several prototypes has been measured with an high intensity pion beam of 350~MeV/c energy at the PSI. 
Figure~\ref{fig:fig18} left shows the detector current as a function of time for particle rate of about 100~kHz/cm$^2$. Some discharges (seen as high current peaks) are visible, more frequently on one of the two SBU-type detectors.
On the  right part of the plot  the spark probability is reported as function of the amplification HV for the PAD-P, DLC-20 and two SBU detectors.
The PAD-P detector shows a very high stability with a spark probability less than 2$\times$10$^{-9}$/pion/cm$^2$. 
DLC20 shows better stability of the two SBU, explained by the lower resistivity of the external resistive layer for the latter.

\subsection{Towards large size detector: experience and future developments}
\begin{figure}
\begin{center}
\includegraphics[width=\linewidth]{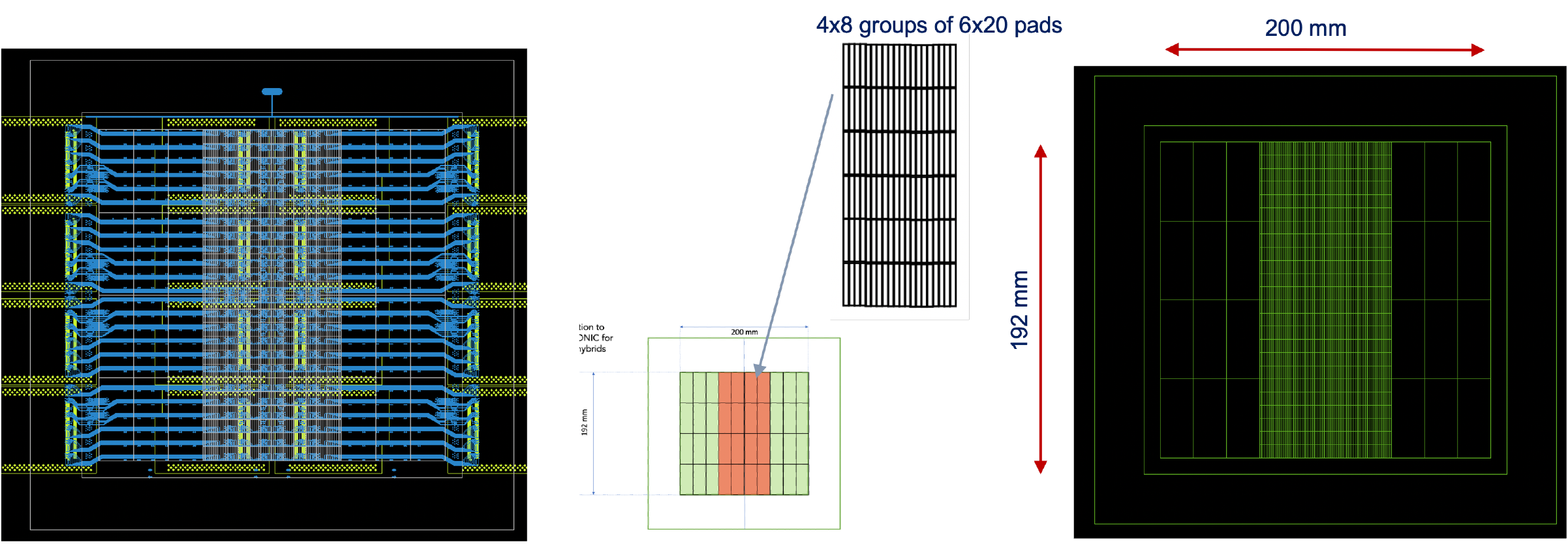}
\end{center}
\caption{Schema of the large-size demonstrator under construction. In the center is also shown the layout of the readout elements.}
\label{fig:fig19}
\end{figure}

\begin{figure}
\begin{center}
\includegraphics[width=\linewidth]{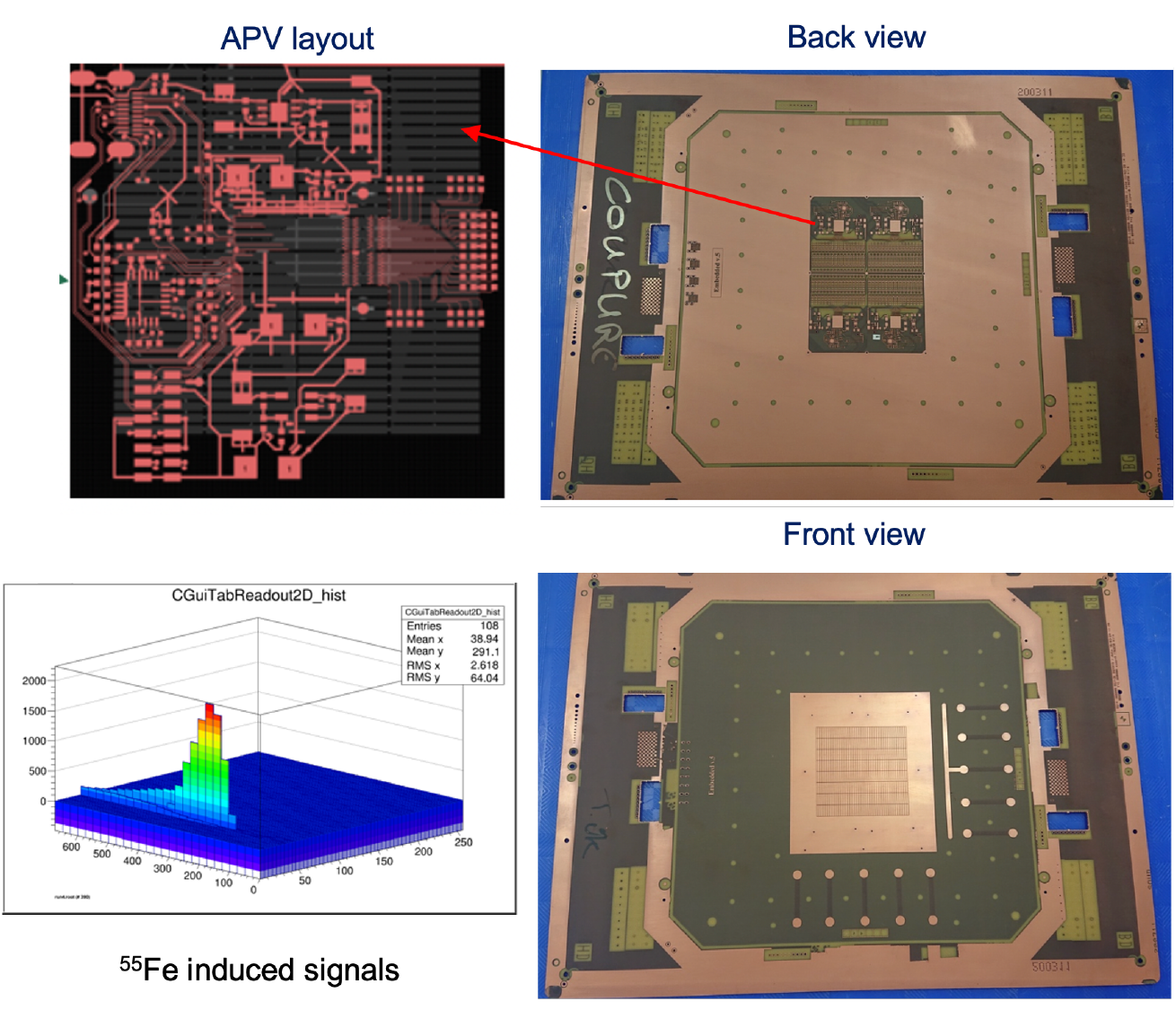}
\end{center}
\caption{Pictures of the first implementation of the PAD detector with integrated electronics. On bottom left part 
a signal induced by a photon form a $^{55}$Fe source acquired with the PAD detector with integrated electronics is reported.}
\label{fig:fig20}
\end{figure}

The detectors presented in the paper fulfill the requirements on rate capability needed by a gaseous detector aiming to be used  in high particle rates environments 
where very good tracking performance and reliable operations are needed.
Anyway, the studies have been  so far conducted on detectors with a limited active surface, of the order of few tens of cm$^2$.  
To demonstrate the feasibility of building detectors with larger surfaces that maintain the same performances and stability, we are building a detector 
with an active area of 192$\times$200~mm$^2$ with 4800 readout pads with a pitch of 1$\times$8~mm$^2$ based on the SBU technology. The schematic layout of this larger prototype is shown in Figure~\ref{fig:fig19}.
Anyway, the major difficulty to scale our high-granularity detector to larger dimensions consists in a complex and difficult routing of each individual pad signals to the readout electronics hosted on its border.
In order to obtain a detector scalable to larger sizes, we are studying the possibility to bond the readout electronics on the back of the detector PCB, 
occupying not more than the space of the active area, thus reducing the inactive surface of the detector.
An "embedded electronics" prototype based on the APV25 chip has been produced at the CERN MPT workshop.
It has a total active surface of 64$\times$64~mm$^2$ with 512 pads large 1$\times$8~mm$^2$. matching the APV25 chip channels capability and dimensions.
Fig.~\ref{fig:fig20} shows front and back views of the PCB with the chips associated circuitry,  We have performed first tests with APV chips, obtaining encouraging results. 
Anyway, the work on this direction is still in progress. After the electronics integration, our plans are to integrate in the detector PBC also a cooling system if chips requiring continuous cooling has to be used. 
The idea here is to include a micro-channel cooling loop inside the base-plane material of the detector 
to have a compact and highly integrated detector system which includes the sensitive device and the electronics with its cooling.

%

%

\subsection{Conclusions and final remarks}
Since 2015 our R\&D team is working on the development of a new generation of single amplification stage resistive Micromegas suitable for applications at high particles rates.
This paper report a summary of the performances obtained by several detectors that have been built with different construction techniques, having in common the segmentation of the readout elements, with a density of more than 30 
readout pads  cm$^2$. All the detectors have been fully characterized with tests in laboratory and at beam lines: Results show that they can be efficiently operated up to particle fluxes of 10~MHz/cm$^2$, 
have an efficiency to charged particles above 98\% and a spatial resolution ranging between 80 and 200~$\mu$m (depending on the technology) for readout pads with 1~mm wide pitch.
Future steps of this R\&D project will be undertaken towards the construction of large-size detectors and the development a fully integrated system that will include the the front-end electronics and the cooling in a single structure; 
Studies conducted so far show that it's possible to choose the resistive layout, the readout segmentation and the operating conditions in order to optimize the detector performance for the specific application.

\subsection{Acknowledgements}
We gratefully acknowledge the CERN MPT workshop (in particular R. de Oliveira and his group, for ideas, discussions and the construction of the detectors), E. Olivieri and the RD51 Collaboration for support with the tests at the Gas 
Detector Development (GDD) Laboratory and for the test-beam at CERN. The team of the piM1 Beam facility at PSI  and GIF++ facility team also deserve our full and heartfelt gratitude for their support during our test beam and irradiation activities.

We would like to thank the organizers of Snowmass 2021 (US Community Study on the Future of Particle Physics) and in particular the conveners of the white paper 'Recent advances and Current R\&D'.

\section{Trigger extensions for the scalable readout system SRS-e}\label{sec-5}
\subsection{SRS-e extended Scalable Readout System}
SRS is a widely used readout system~\cite{Mar11} for low-to high channel-count, gas or photon detectors of type MPGD  or SiPM~\cite{Chr18}. ``SRS classic'' was designed for scalability from desktop systems to rack-sized readout systems running under the same Online DAQ and Control system\cite{Est16}. The new, extended SRS-e paradigm adds realtime trigger functionality, deep trigger pipelines and a generalized frontend link via the new eFEC concentrator cards which form the SRSe readout backend. Horizontal links synchronize clocks and realtime actions, and the vertical links can be connected within vertical or horizontal readout architectures. Compared to SRS classic, where crate-based FEC cards concentrate 8 frontend links each, the  eFEC doubles the frontend links to 16, enhances the output bandwidth to 20\,Gbps and includes an 80W subsystem to power the frontend. A single eFEC concentrates 16 links to readout ASICs which are integrated on hybrid frontend carriers, generally named hybrids. Large SRSe systems are stacks of eFEC’s powered in SRS crates connected to detector-resident hybrids. The default vertical link protocol is DTCC for Data, Trigger, Clock and Controls, implemented over physical HDMI cables for transmission of 32 or 64-channels per ASIC over 2 or 4 LVDS links. An Ultrascale+ FPGA in the eFEC decodes and stores up to 45\,\textmu s of event data in its embedded 18\,kbit BRAM pipelines, allowing in parallel for realtime trigger algorithms to qualify event significances and to format events before transmission to two output links per eFEC. Firmware and DAQ software will get bootstrapped from the classic SRS system which is in full production and has been fully tested up to high rates and multichannel testbeam testups~\cite{Pfe22,Sch20} with the default VMM3a~\cite{Iak19} frontend hybrids. Open for integration of a wider variety of upcoming frontend technologies~\cite{TopFE21}, the generalized eFEC frontend interface can be configured via firmware to operate with readout protocols at up to 3\,Gbps over 4 $\times$ LVDS lines.
\subsection{eFEC extended Frontend Concentrator}
e-FEC is an acronym for extended Front-End Concentrator. Apart from the SRS crates and optional CTF clock fanout, eFECs are the only modules required for an SRSe  readout backend. The Online system gets connected via network fibers or cables with two 10\,GBE uplinks per eFEC.  The choice of the frontend ASICs is in principle open as long as the corresponding readout data is transmitted in digital format over LVDS links using a protocol like 8b10b. Adapter cards for different types of frontends are not required any more due to the configurable digital link interface. The main functional extension compared to the classic FEC is the realtime trigger functionality which can now get implemented in firmware on the Ultrascale+ FPGA, making use of a vast amount of DSP and programmable logic resources. The trigger algorithms can proceed in parallel to the data pipeline transit time with realtime access to event markers and delimiters. With the default VMM3a fronted~\cite{Lup18}, 2k channels are transiting the eFEC pipeline before transmission to the Online system. Firmware will be bootstrapped from the production firmware developed for `Assister cards' of ESS, and accessible in an open-source repository under the condition that newly developed user-firmware is added to the same repository. The data-driven paradigm of VMM3a-based experiments will be enhanced by use of realtime ART flags (Address in Real Time)~\cite{05-ATLASNSW} as event-framer. Experiments which require an externally generated trigger can connect prompt triggers in the same way as on the classic FEC via a NIM pulse. The eFEC however calls for additional detector-specific firmware algorithms to be developed for detection and selection of event signatures in time and space, or by generation of multiplane-track triggers from hit address combinations. Further examples for online triggers are total energy sums, or charge-over-threshold in space-time regions. Such triggers require extension of the classic SRS slow controls via dedicated Gigabit network links connecting to the Online and Control system. The addition of realtime-programmable trigger primitives is on the shelf for project-development by teams, making use of the FPGA-embedded dual-core realtime subunit, which may further get extended to a machine-learning trigger concept. Trigger boundaries (2k channels for VMM3a frontends) can be overcome via the horizontal X-link bus connected between eEFC frontpanels, connecting neighboring eFEC's in bidirectional, high-bandwidth ring topologies for common trigger regions. The second important extension of SRSe is a generalized frontend link interface, defaulting to the SRS-classic DTCC protocol~\cite{VMMFrontend} used for readout of VMM3a hybrids, or the SiPM frontend of the NEXT experiment. The new link interface offers 4 individually selectable link directions and firmware-defined protocols. SRSe maintains HDMI link ports on the eFEC rear side, each interfaced via 3\,Gbps LVDS Rx-TX macros with directional configuration. 
\subsection{SRS-e extended readout architectures}
SRS classic was designed as scalable architecture via incremental addition of independent vertical readout slices, from frontend to backend allowing to bootstrap systems from a minimal number of test channels to full size.  SRSe adds the cross-boarder X-link to the hierarchical architecture and thereby offers the hierarchical architecture as alternative.  Due to parallelism in the vertical slice architecture, the total bandwidth can achieve very high levels, only limited by the network capabilities of a  network switch which transmits the concentrated data from 16 frontend links of a single eFEC to online computer(s).  Overload of the network can occur in a case of mismatch between the aggregate output data bandwidth and the capacity of the switch or  input bandwidth of the online system. Such overloads are mitigated by the use of hardware zero-suppression  in the frontend ASICs and  implementation of realtime triggers in the eFECs. Horizontal X-link trigger topologies further help to optimize such triggers.  The horizontal S-bus serves for common, simultaneous actions at the 25\,ns level, like Busy, Xon/Xoff, or Reset. Analogue lines of S-bus allow to synchronize with external analogue conditions like spill structures or B-fields. The CTF-e common clock module is required for the use of multiple eFECs. CTF-e is an extended `CTF classic' module, synchronizing the eFEC system clocks with either a common external clock source, or an  eFEC  used as master clock generator. The clock can be either a local CTF-e clock or an external clock, like the master clock of an accelerator.  

The hierarchical architecture option of SRSe reproduces implementations like the ALICE EMCal hierarchical trigger~\cite{Mul10} with trigger levels in successive FPGA layers at increasing trigger latencies, requiring additional data buffering at the higher levels.  For this purpose, the eFEC can be equipped with DDR4 plugin memory plugin module of up to 64 Gbyte which is directly interfaced to the 64\,bit bus of Ultrascale+ FPGA. The output bandwidth of the hierarchical architecture is limited to 2 $\times$ 10\,Gbps.  The X-link, S-bus and CTF clocks links are connected in the same way as for the vertical slice architecture.   
\subsection{eFEC control and test subsystem}
eFEC cards can be equipped with a 32\,bit SoC plugin mezzanine, including USB and wireless connectivity to external computers. Though the FPGA-embedded 4-core Cortex CPU  has  Monitoring access to all hardware resources, the external SoC card with its embedded uPython libraries and flash-resident memories offers a user-friendly development and test environment via a terminal line to a Laptop for an initial test and debugging period. The SoC mezzanine is used in similar ways as on the VTC tester~\cite{VTC} for VMM production testing, where I/O control levels, I2C and SPI buses are routed to all programmable resources and a comfortable user-interface with parameter storage to a database on a Laptop. The eFEC is also connected to programmable resources in the frontend. In the simplest case these are configuration settings for frontend ASICs oven an I2C bus  mastered from the eFEC.  Examples for local programmable resources on eFEC are  supply voltage settings, current-monitoring and configuration of Flash devices. Resources on the frontend include ASIC configurations, baseline and pulser test monitoring and readout of sensors like geographical position, temperature, humidity, B-field or orientation. The SRSe control system, like the existing online control system of SRS will provide full control over all resources in backend and frontend and return key parameters like link status, rates, baselines, temperatures, geographical positions and so forth. The SoC plugin  gives access to the same resources at a register level, for sanity checks, test of  special settings and new firmware. 
\subsection{eFEC subsystem for frontend power}
Digital ASICs integrated on frontend cards consume between 10 and 20\,mW per channel and require supply voltages between 0.8 to 1.2 Volt. This does however not include the power-loss over cables, drop-off in voltage regulators and power for additional integrated logic like FPGAs, Flash devices, link drivers, sensors, monitoring ADCs and bias generators. The SRS default readout frontend, the VMM3a hybrid for SRS with 128 channels requires 2 supply Voltages for two ASICs and all support chips on the ASIC carrier card, totaling to : 3\,V@0.2\,A and 2\,V@1.7\,A roughly 4\,W per 128 channels, or effectively 30\,mW/channel compared to the 12\,mW specified for the VMM ASIC only. As a rule of thumb, the power budget for the a readout frontend should be dimensioned with a factor of 2 of the power/channel-specification of the ASICs and a secondary supply voltage needs to be added for all service and monitoring logic. In order to provide such a general-purpose power interface for the SRSe frontend, eFECs implement a programmable 80~Watt, two-voltage subsystem. The primary supply for ASICs requires a high-current programmable supply of $\cal{O}$(1.8\,V) for frontend-resident LDO’s and the service logic requires normal-level programmable supply of $\cal{O}$(3.3\,V). These nominal voltages can be programmed in a range with current monitoring and auto-fusing. With 90\,\% efficient DC-DC conversion, the maximal delivered 80~Watt, the eFEC produces $\cal{O}$(8\,W) of heat for the crate cooling. The dual power panel connector on the eFEC frontpanel assumes the use of the PMX Voltage distributor box~\cite{PMX} with external Volt/Ampere monitoring per ASIC card and including a high-current, low impedance return GND path to the crates. This power distribution concept was validated in testbeams with a multi-FEC VMM frontend and which is generally recommended to avoid ground lifts and/or redundant currents flowing across the detector Ground. 
\subsection{Crate environment}
The eFEC card format is 6U $\times$ 220\,mm for insertion in rack-mountable subracks which comply to DIN~41494 and IEC~60297-3. Crates are mechanically compatible with the ``SRS classic'' format however may require upgrade of the ATX power supplies because eFECs get exclusively powered via SATA power cables delivering per eFEC up to 110\,W of which up to 80\,W are dedicated to the external frontend. Up to 4 eFECs can be powered from 600\,W ATX power supplies of type ``80PLUS'' which conform to more than 80\,\% power conversion efficiency from universal AC inputs of 110-240\,V. The legacy ATX adapter~\cite{ATX} modules used in classic SRS crates are not required for powering eFECs. Minicrates of size 3U $\times$ 220\,mm have one ATX power supply for powering up to 2 horizontally inserted, eFEC cards and one CTF clock card. Connected to VMM frontends, up to 32 HMDI port links from 2 eFECs add up to a 4k channels for a fully-loaded SRSe Minicrate with 4 uplinks of 10\,Gbit each. Eurocrates of size 6U x 220\,mm have two ATX power supplies for up to 8 vertically mounted eFECs and one CTFe clock card. Connected to a VMM frontend, a Eurocrate can concentrate up to 128 HMDI link ports from up to 8 eFECs, adding  up to a 16k channels with 16 uplinks of 10\,Gbit each. Readout backends beyond the channel-capacity of single crates can be incrementally increased, preferable in rack-mounted subracks with interleaved ventilators below each Eurocrate or below every 2 Minicrates. A total of more than 8 FECs requires addition of one more CTF card per octal eFEC group, all  driven from a common external clock source. A portable eCLK box, powered from the Crate 5V power outlet, will become available for driving multiple CTF cards with a common, jitter-free clock  or  an external clock. Any mixture of Eurocrates and Minicrates is equally possible. Due to the accumulation of the return currents (up to 17\,A per octal group of 8 VMM hybrids), a very low-impedance ground return copper braid fanout should be connected to the rear-side GND attachment plates of the crates. With 1 copper braid per group of 8 VMM carrier-cards, 2 copper braids are needed per Minicrate, or 8 per Eurocrate.      
\subsection{Status of SRSe with the VMM3a frontend}
SRS is a very mature readout technology developed since 2009 with resources of the RD51 collaboration and using CERN infrastructure and RD51 resources~\cite{05-RD51}. Many RD51 collaborators have contributed to its progress with wide acceptance within the MPGD user community and beyond. Following a very successful early period with the analogue APV frontend, newer frontend technologies, like SAMPA~\cite{Her20}, Timepix
~\cite{Gru19} and in the particular VMM, have been interfaced to SRS. Based on the latest VMM3a ASIC version developed by BNL for the ATLAS NSW detector, SRS was fully redesigned on all levels for commercial SRS production. By 2021, 250k VMM channels are either in use or  in production for new installations. The SRSe extension represents the next generation of SRS, based on many lessons learnt from deployed SRS classic systems and adding new functionalities and higher performance.  At the time of writing the eFCE module is fully specified from the system level down to the schematics and 3D levels, with delays due to the general ASIC shortage in 2021/22 the first 2 prototype cards are now expected for Q2 2022 whilst firmware developments on the similar Kintex family with the VMM3a fontend are already in production.
\subsection{Work plan}
This LoI represents also a call for competences to implement FPGA-based trigger functionality with GUI-based user control on the new eFEC platform. As soon as eFEC hardware will become available, the default firmware developed for the VMM frontend on classic FECs (Virtex-6) and ESS Assister cards (Kintex-7) will be adapted to the eFEC Ultrascale+ Zync FPGA as baseline for standard readout firmware already existing in the SRS classic systems. The implementing of the new trigger features will be gradual and driven by requirements of the associated experiments.

The initial phase consists in bootstrapping of firmware and software on eFEC prototypes in order to establish a baseline readout system, ready for testing and commissioning with user-defined frontends.

A template work plan includes the following
\begin{itemize}
  \item Use existing frontends (VMM3a, SiPM adapter) as well-known frontend as baseline
  \item Establish a common user forum modelled after RD51 WG5.1 user group to identify priorities and agree on code exchange rules and practices
  \item With first eFEC prototypes, establish a basic set of uPython test procedures for eFEC sanity checks  and register-level access to all connected resources via I2C, SPI and JTAG
  \item Establish a Linux-based control and monitoring environment on the embedded multicore system 
  \item After initial production of 2 pilot eFECs, launch a first batch production 
  \item Deploy first stand-alone eFEC readout systems with VMM in testbeam-like environment
  \item If required, develop a MAC and/or Windows-based GUI for embedded use of uPython register level procedures
  \item Work with new experiments on the implementation of  basic sets of triggers (fast-or, veto, coincidence, region, topology, etc)
  \item Identify experiments requiring different frontend ASICs for implementing other kinds of link interfaces
  \item Establish rules for parameterization of trigger primitives     
  \item Establish a standard user guide for shifters
\end{itemize}

\subsection{Additional Material}
\begin{figure}[!ht]
    \centering
    \includegraphics[width=\textwidth]{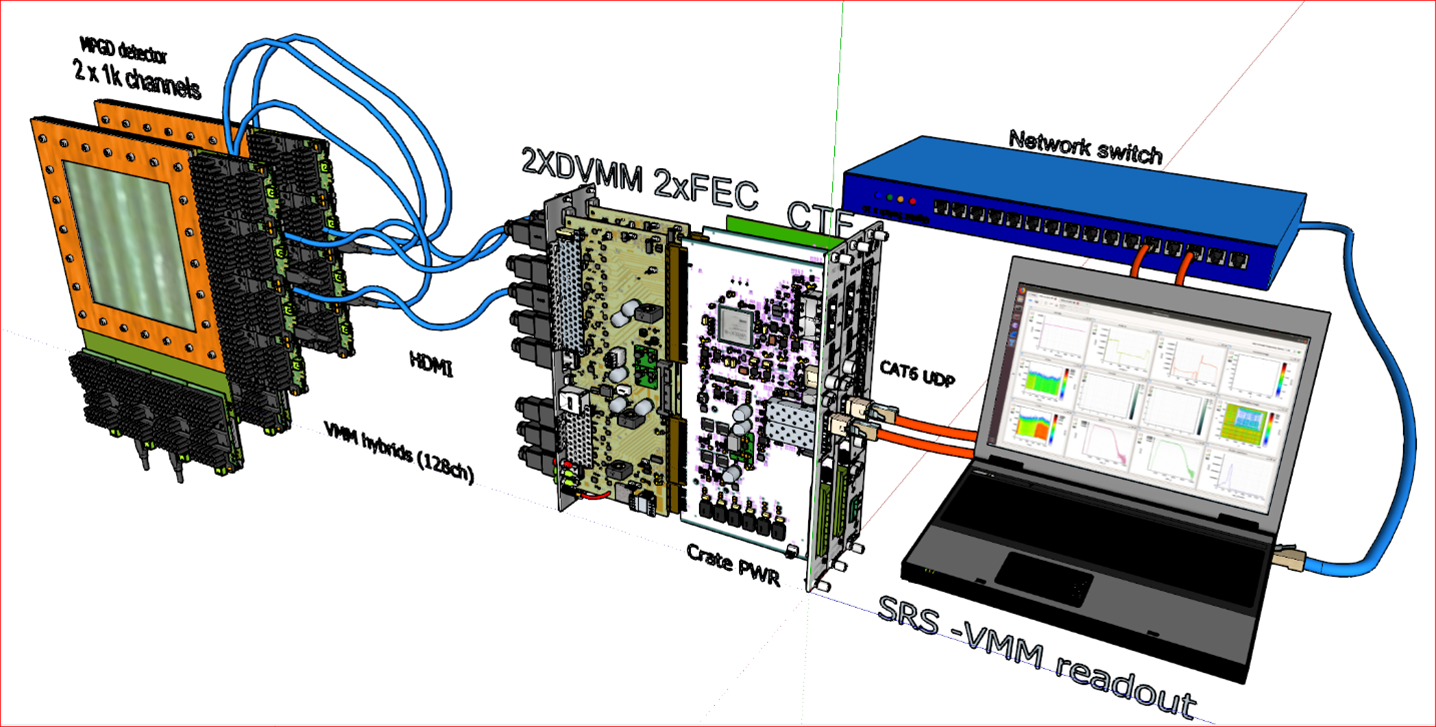}
    \caption{SRS classic small readout systems with FEC, CTF, DVMM adapter, frontend links (HDMI) and VMM3a fronted hybrids attached to 2 GEM planes.}
    \label{fig:04-Picture1}
\end{figure}

\begin{figure}
\centering
\begin{minipage}{.45\textwidth}
  \centering
  \includegraphics[height=8cm]{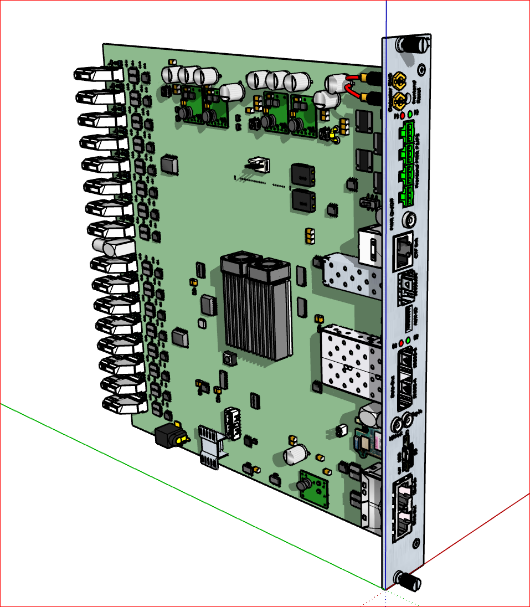}
\end{minipage}%
\begin{minipage}{.55\textwidth}
  \centering
  \includegraphics[height=8cm]{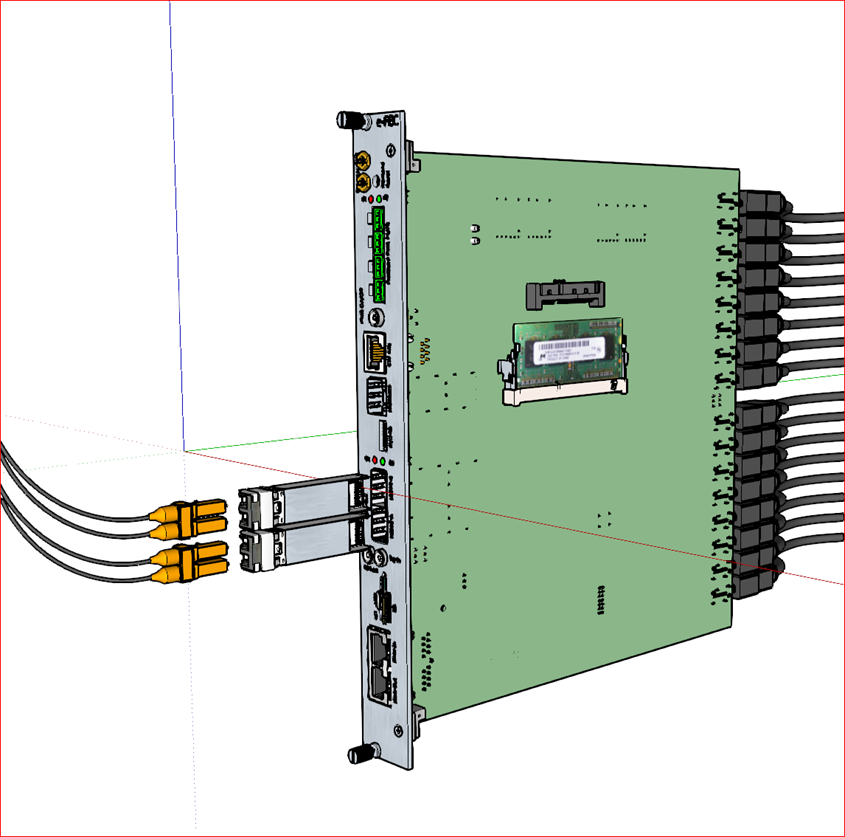}
\end{minipage}
\caption{Left: New eFEC module 6U for SRS crates with cooled Ultrascale+ FPGA and 16 generalized HMI link ports. DCDC converters provide programmable power to the frontend. Right: eFEC with rear-side generalized frontend links and two SFP optical output links. The DDR4 data buffer is an optional plugin.}
\label{fig:04-Picture2-3}
\end{figure}

\begin{figure}
    \centering
    \includegraphics[width=\textwidth]{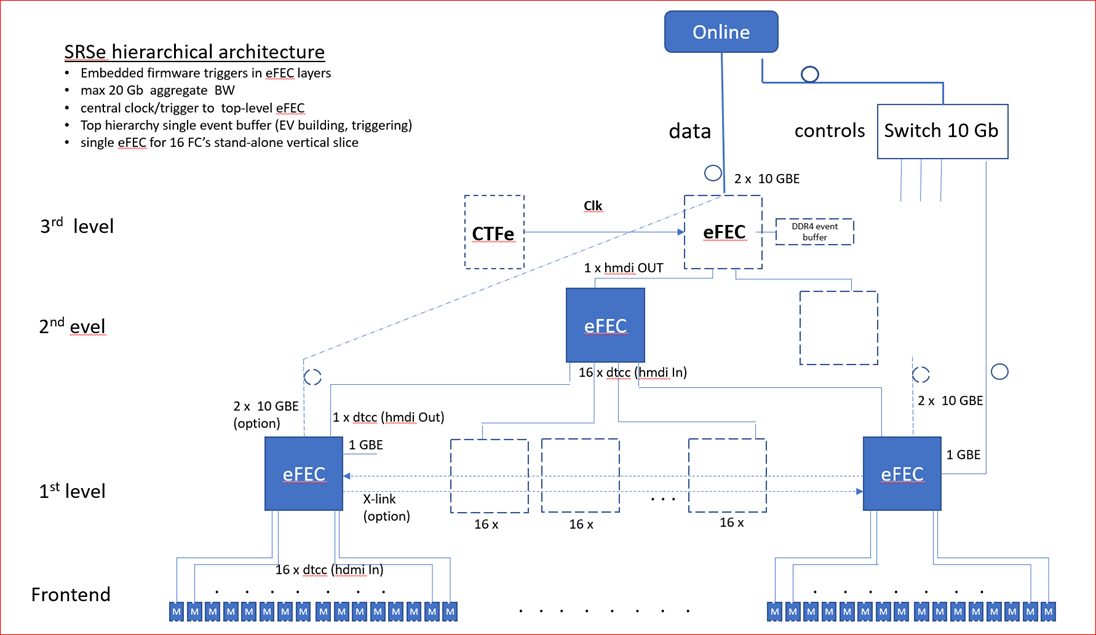}
    \caption{SRSe hierarchical architecture.}
    \label{fig:04-Picture4}
\end{figure}

\begin{figure}
    \centering
    \includegraphics[width=\textwidth]{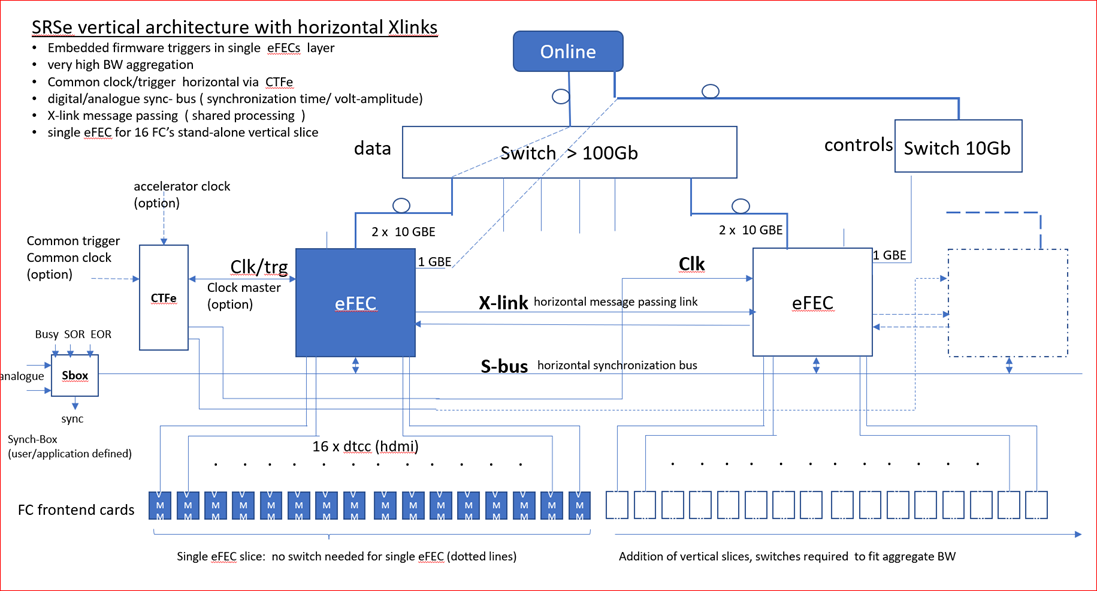}
    \caption{SRSe vertical slice architecture.}
    \label{fig:04-Picture5}
\end{figure}

\begin{figure}
    \centering
    \includegraphics[width=0.75\linewidth]{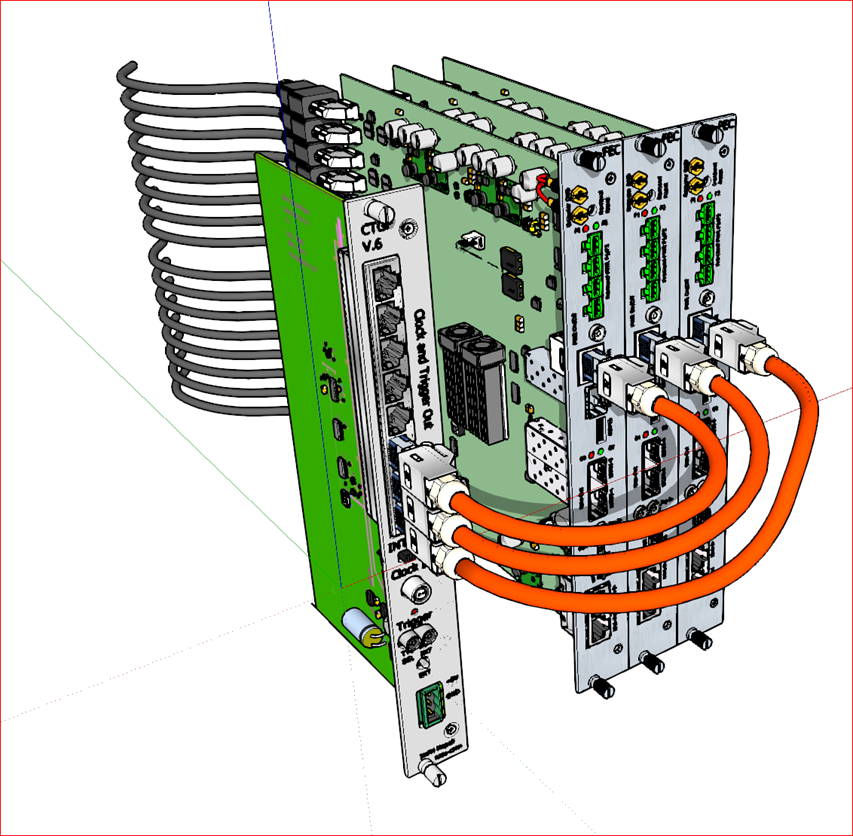}
    \caption{Common clock and trigger for stack of eFECs.}
    \label{fig:04-Picture6}
\end{figure}

\begin{figure}
    \centering
    \includegraphics[width=\textwidth]{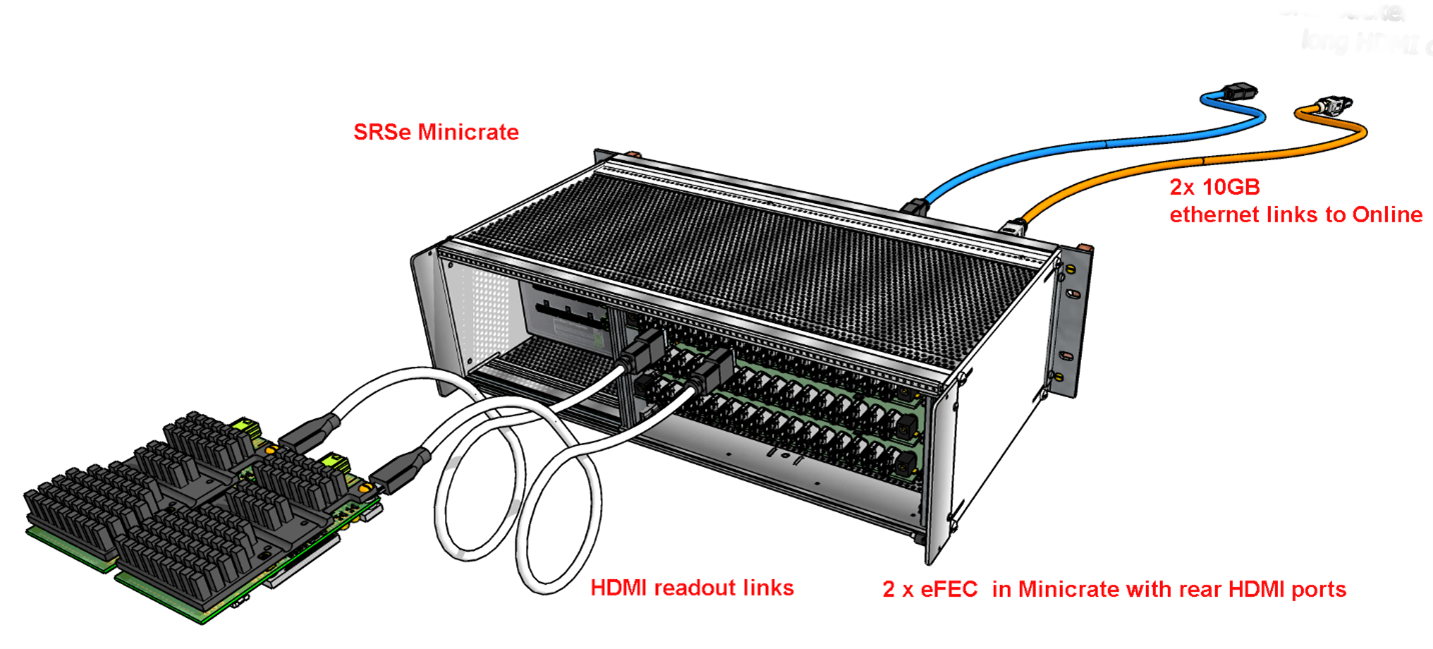}
    \caption{2 $\times$ eFECs with two HDMI readout links for 2 VMM3a hybrids. eFECs housed in a simplified 3U Minicrate (SATA power only).}
    \label{fig:04-Picture7}
\end{figure}

\begin{figure}
    \centering
    \includegraphics[width=0.75\textwidth]{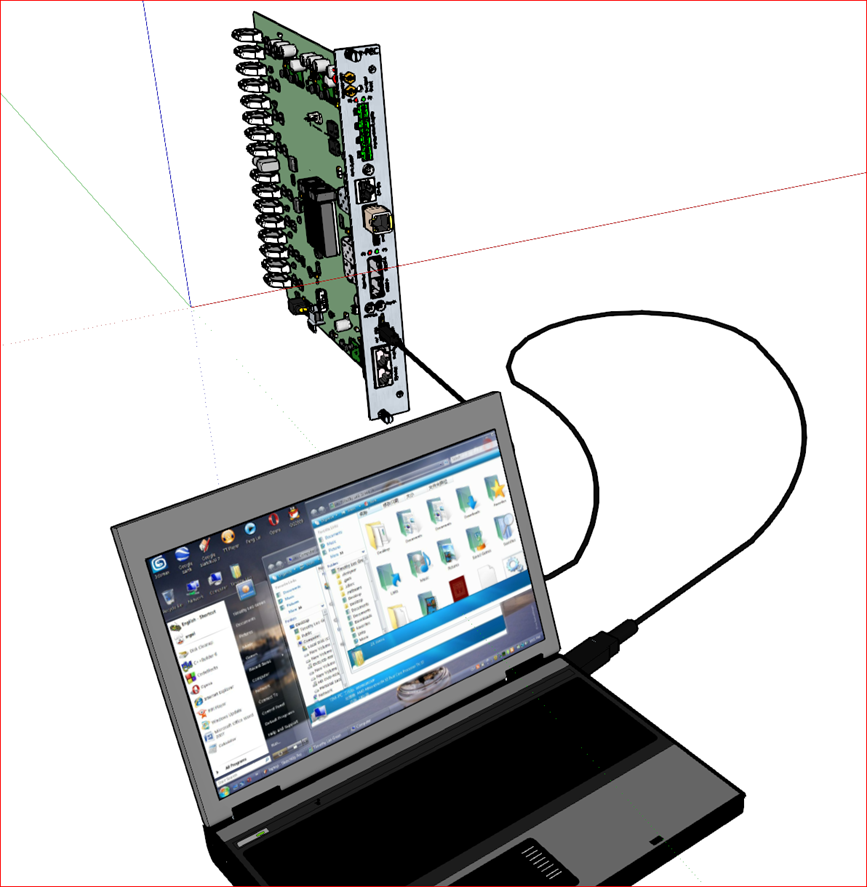}
    \caption{Local eFEC configuration and tests via USB link.}
    \label{fig:04-Picture8}
\end{figure}

\begin{figure}
    \centering
    \includegraphics[width=\textwidth]{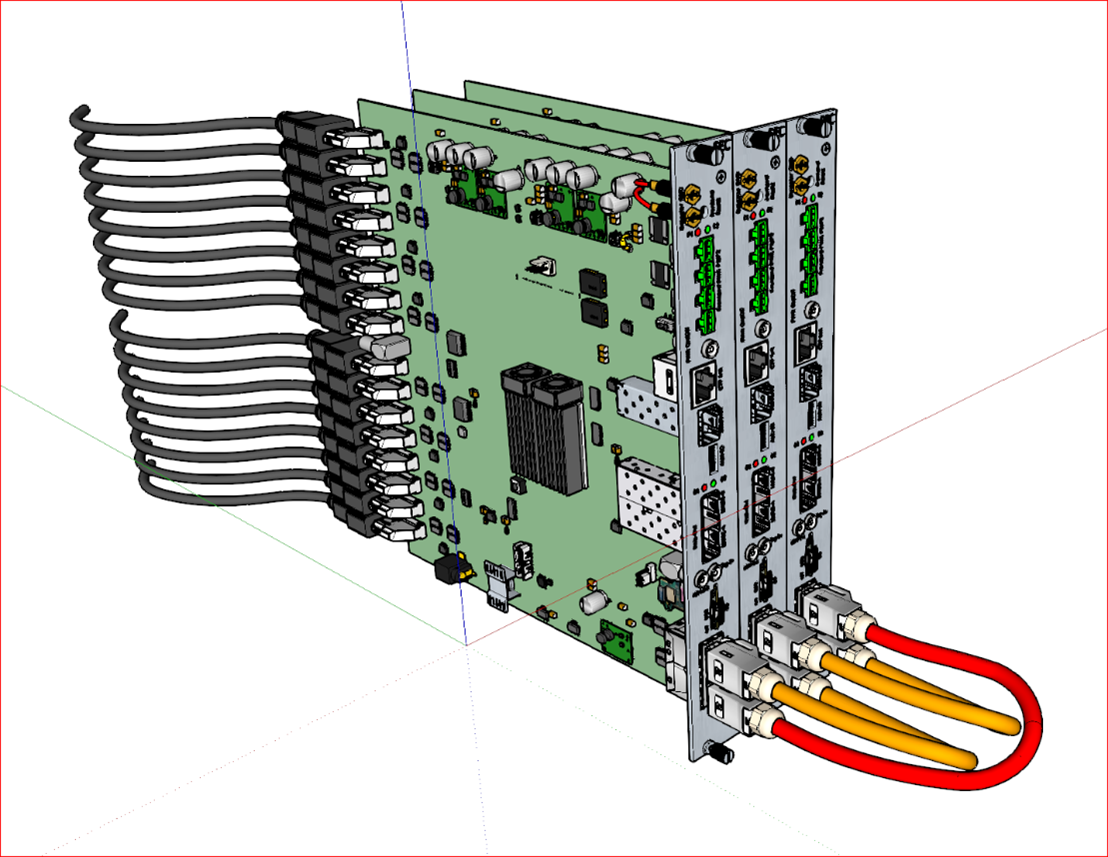}
    \caption{Example of a regional ring topology connected via Xbus network cables.}
    \label{fig:04-Picture9}
\end{figure}

In the following one can find pictures (\Cref{fig:04-Picture1,fig:04-Picture2-3,fig:04-Picture4,fig:04-Picture5,fig:04-Picture6,fig:04-Picture7,fig:04-Picture8,fig:04-Picture9}) of the items discussed in the previous sections.

\section{High-gain, low ion-backflow double and triple micro-mesh gaseous structures}
\subsection{Motivation}
Positive Ions produced by electron-molecule collisions during the avalanche process can flow back to the drift region and affect the detector performance in many ways. In gas photon detectors, the ion bombardment can reduce the quantum efficiency of the photocathode or even render it completely ineffective. In the high-rate time projection chambers (TPC), large number of ions flowing back from the amplification region will cause significant non-uniformity of the local electric field in the drift volume and eventually lead to a distortion of the charged particle tracks. 
Micro-pattern gaseous detectors with high gain and very low ion-backflow (IBF) provide cost-effective solutions to large-area and position-sensitive photon detection and readouts of high-rate TPCs. The backflow fraction of Micro Pattern Gaseous Detectors (MPGD) is smaller than that of the more classical configuration. For example, in the Micromegas detector, a large fraction of the secondary positive ions created in the avalanche can be stopped at the micro-mesh depending on the field ratio, detector geometry, etc.
\subsubsection{High rate TPC in future Colliders}
\begin{figure}
 \centering
 \includegraphics [width=4.29167in,height=1.20833in]{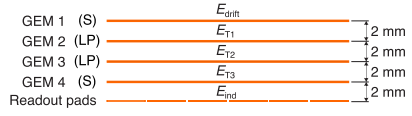}
 \caption{Schematic setup with four GEMs for ALICE TPC \cite{citation-1}.}
 \label{fig:05-Picture1}
\end{figure}

It is worth mentioning that the upgrade of the large TPC of the ALICE detector at the CERN LHC provides a successful example of high rate TPC in the real experiment. In the ALICE TPC, The requirement to keep the ion-induced space-charge distortions at a tolerable level leads to an upper limit of 2\% for the fractional ion backflow at the operational gas gain of 2000 \cite{citation-1, citation-2, citation-3}, which was achieved by using stacks of four GEMs, as shown in Fig.~\ref{fig:05-Picture1}, GEM foils with standard (S, 140 $\mu$m) and large (LP, 280 $\mu$m) hole pitch are combined to an S-LP-LP-S configuration. As well as the electric field was also highly optimized with a very low transfer field E$_{T3}$ between GEM 3 and GEM 4 of only 100 V/cm, whereas the other transfer fields and the induction field E$_{ind}$ are kept at typical values around 3500 V/cm.
However, the method of the ALICE TPC with four GEMs may be difficult to apply in future large colliders with higher event rate. The International Linear Collider (ILC) is a proposed electron-positron collider for Higgs precision measurements and discovery. The TPC scheme is conceived to be central tracker of the International Large Detector (ILD) concept for the ILC, in which, the IBF ratio should be smaller than 0.1\% at an expected gain of $\sim$ 5000. It is like the Circular Electron Positron Collider (CEPC), its TPC solution also requires very IBF ratio at 0.1\% level \cite{citation-4, citation-5, citation-5a}.
\subsubsection{Gaseous photon detection}
Gaseous photon detector (GPD) using micro-pattern gas detectors (MPGDs) have been widely studied owing to their potential advantages, such as large effective area with low cost, high spatial and time resolutions, and high magnetic field resistance. The most common examples are photon detectors for ring-imaging Cherenkov counters (RICH) \cite{citation-6, citation-7} and gas photomultiplier tubes (gas-PMTs) \cite{citation-8,citation-9,citation-10}. However, the typical gas-PMT gain is $\sim 10^4$ whereas regular vacuum PMTs have a gain of $\sim 10^6$. Another big challenge in application of gas-PMTs is for IBF suppression. The effect of ion-backflow may cause aging of the photocathodes of GPDs. The quantum efficiency of a GPD will degrade when its photocathode is bombarded by ions \cite{citation-11, citation-12}, which are produced during the gas multiplication process. Some detector structures, based on micro-pattern gaseous detectors (MPGD), have been studied \cite{citation-13} to suppress the IBF ratio, such as the multiple gas electron multiplier structures (multi-GEMs) \cite{citation-14}, hybrid structure \cite{citation-5}, cascaded GEM to micro-mesh gaseous structure (Micromegas) \cite{citation-15}, and micro-hole and strip plates (MHSP) \cite{citation-10, citation-16}.The studies mentioned suggest that mesh-type MPGDs have a better IBF suppression capability than hole-type ones. This therefore provides motivation to fully explore the mesh-type structure for gaseous photon detection.

\subsection{Concepts and R\&D status}
\subsubsection{Double micro-mesh gaseous structure (DMM)}
\begin{figure}
 \centering
 \includegraphics[width=3.56262in,height=1.45833in]{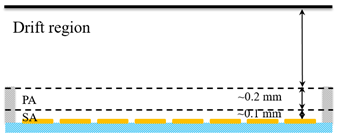}
 \caption{Schematic design of the DMM.}
 \label{fig:05-Picture2}
\end{figure}

\begin{figure}
 \centering
 \includegraphics[width=0.65\linewidth]{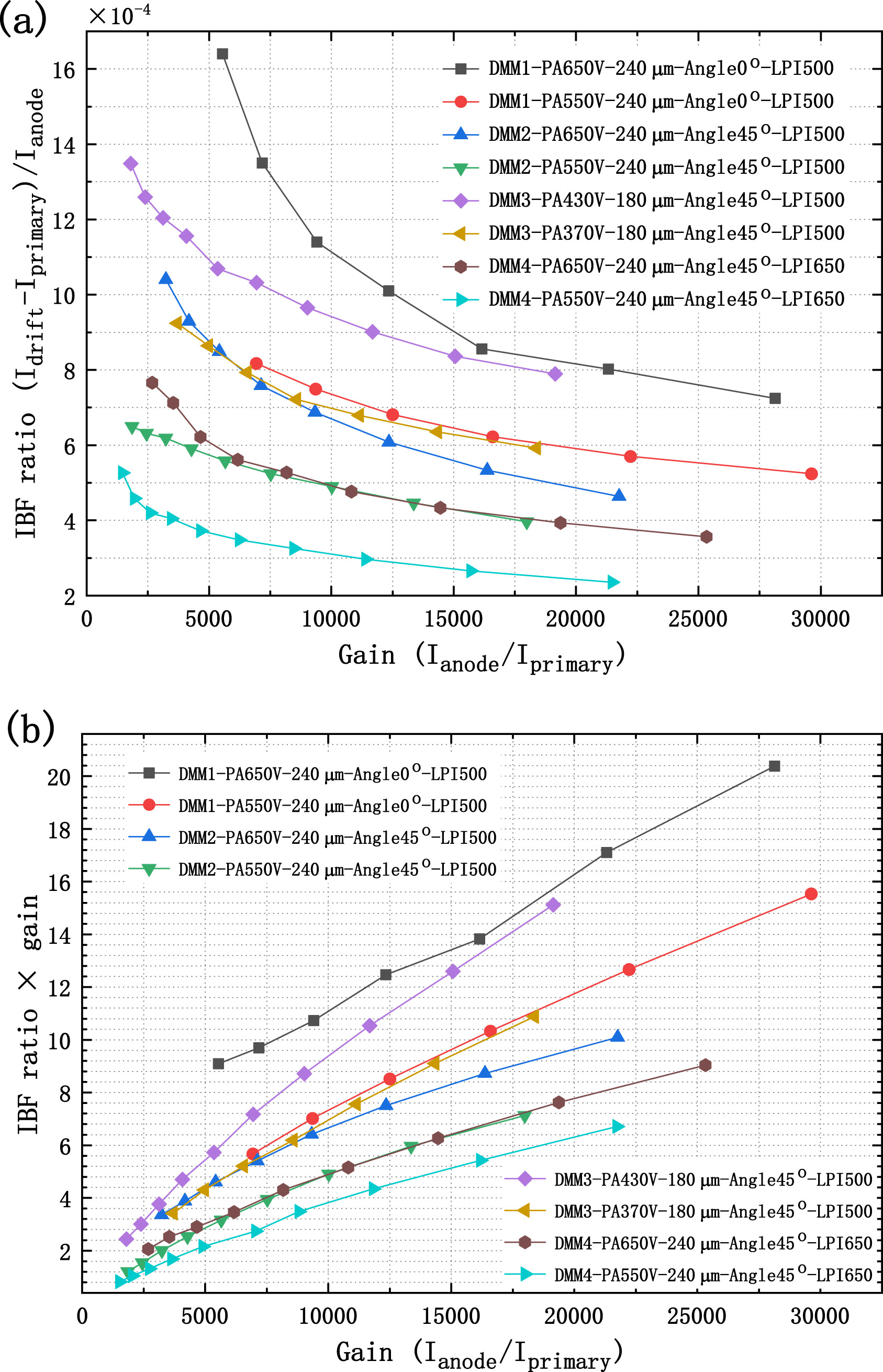}
 \caption{(a) IBF ratio versus total gain with a fixed PA voltage, for all detectors; (b) the product of the IBF ratio and gain versus the total gain.}
 \label{fig:05-Picture3_4}
\end{figure}

For the purpose of high gain and low IBF, a double micro-mesh gaseous structure (DMM) \cite{citation-17, citation-18, citation-19} with two avalanche stages, was developed using a thermal bonding method \cite{citation-20}. The structure of a DMM is illustrated in Fig.~\ref{fig:05-Picture2}. It has a 3-5 mm gas gap for particle primary ionization and electron drift, followed by a $\sim$ 0.2 mm PA gas gap and a $\sim$ 0.1 mm SA gas gap that are defined by two meshes and an anode. The structure is quite similar to that of a typical Micromegas except that it has two layers of meshes to provide cascading avalanche amplification. A typical Micromegas has only one layer, hence giving single avalanche amplification. The double cascading avalanche gaps ensure a very high gain for a single electron and, with the proper configuration of electric field, a low IBF ratio. It also preserves the advantages inherited from the typical Micromegas in terms of high-rate capability, good time resolution, and excellent spatial resolution.
The DMM prototypes were fabricated and studied with X-ray and UV lights at first step, its features of high gain of $> 10^6$ and low IBF ratio of $\sim$ 0.05\% was validated \cite{citation-17}. Then the DMM structure was optimized to further suppress its IBF by changing the size of gas gaps and the density of the wire mesh, and more significantly, by aligning the two mesh layers with a crossing angle. A series of DMM prototypes with differing crossing angles, PA gaps, and mesh types were fabricated and investigated using X-rays (5.9 keV from $^{55}$Fe source and 8.0 keV from an X-ray gun). The results (as shown in Fig.~\ref{fig:05-Picture3_4}) of this investigation indicate that a low PA voltage, large PA gap, high mesh density, and crossed mesh setting improves IBF suppression at the same total gain. A low IBF ratio, as low as 0.025\%, was obtained with the prototype, which was made of 650 LPI mesh and had a PA gap of 240 $\mu$m and a crossing angle of 45$^\circ$. In addition, the measurement method for the gas gains and currents was validated. The ion space charge effect was studied and verified to be negligible in the case of very low IBF ratio in a DMM. The stability of the DMM prototype was also measured with a low sparking probability, smaller than $10^{-9}$, in a test done over 20 hrs \cite{citation-18,citation-19}.
\subsubsection{Triple micro-mesh gaseous structure (TMM)}
\begin{figure}
 \centering
 \includegraphics[width=3.63724in,height=1.66667in]{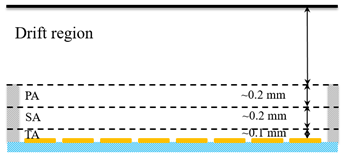}
 \caption{Schematic design of the TMM.}
 \label{fig:05-Picture5}
\end{figure}

\begin{figure}
 \centering
\includegraphics[width=0.75\linewidth]{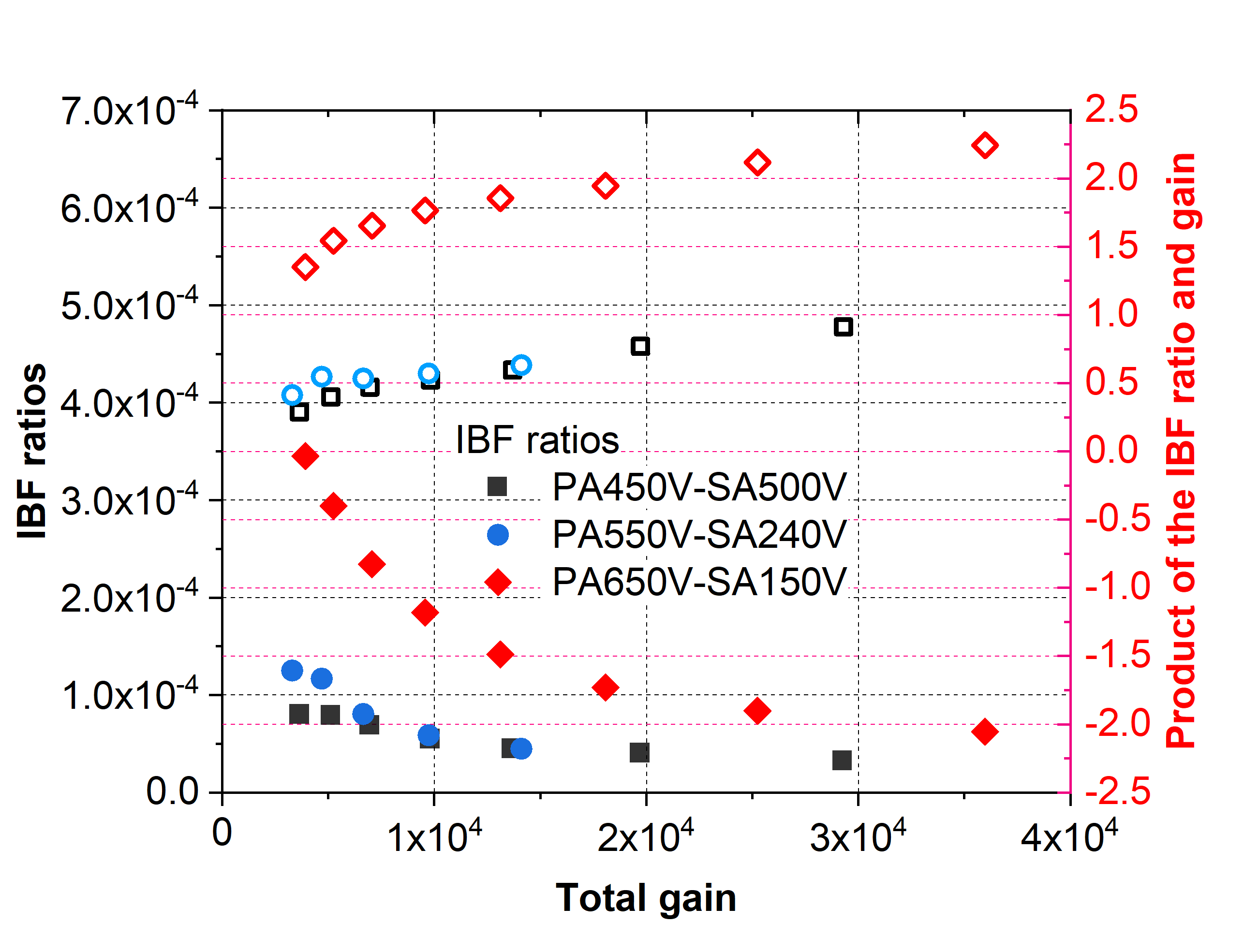}
 \caption{IBF ratio (solid dots) and the product of the IBF ratio and gain (hollow dots) versus total gain with fixed PA and SA voltages.}
 \label{fig:05-Picture6}
\end{figure}

A triple micro-mesh gaseous structure (TMM) is a natural extension to further suppress the IBF that comes from the SA stage of the DMM. 
Following the idea of multi-mesh, a triple micro-mesh gaseous structure (TMM) is a natural extension to further suppress the IBF that comes from the SA stage of the DMM. Fig.~\ref{fig:05-Picture5} shows the schematic design of the TMM, it has one more mesh above the DMM. A TMM prototype was also fabricated with the thermal bonding method \cite{citation-20} and studied to explore the limit for IBF suppression. As shown in Fig.~\ref{fig:05-Picture6} very excellent high gain and ultra-low IBF of $\sim$ 0.003\% is achieved, which reaches one order of magnitude lower than that of the DMM, the lowest IBF value ever obtained with gaseous detector.
\subsection{Plans and objectives}
Following the R\&D works mentioned above, more simulations and tests should be carried out to comprehend the mechanisms and influence factors on the main features of gain, IBF suppression, resolution, etc., to provide guidance for further optimization of the DMM/TMM. We also must notice that the IBF is not the only parameter that decides the choice of the detector, in the previous study, the IBF was optimized while maintaining the others acceptable. Trade-off optimization in different parameters is required for the specific applications. 
\subsubsection{Optimization and improvement}
\begin{figure}[ht!]
 \centering
 \includegraphics[width=0.75\linewidth]{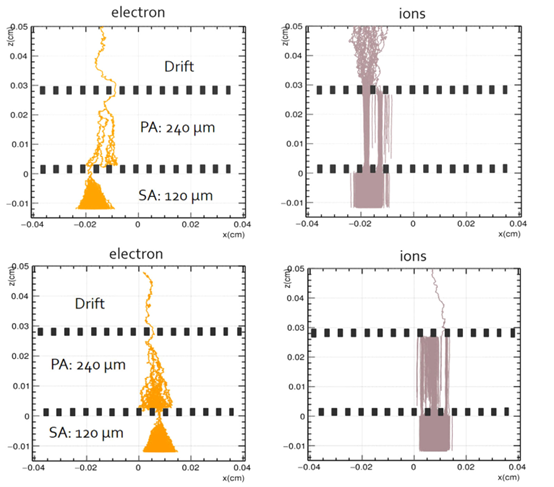}
 \caption{The simulation results of electron avalanche and charge drifting in a DMM structure. The upper and bottom plots show results of two conditions: mesh holes strictly aligned vs. stagger aligned.}
 \label{fig:05-Picture7}
\end{figure}

For the DMM structure, the simulation results as shown in Fig.~\ref{fig:05-Picture7} indicate that excellent low IBF ratio of $\sim 10^{-5}$ can be obtained by strictly misaligning the wires of the micro-mesh, while in the aligned case one can only obtain an IBF ratio of $\sim 10^{-3}$. In the DMM, a low IBF ratio of 0.025\% was obtained by aligning the two mesh layers with a crossing angle, which is a compromise solution to tackling the challenge of strict misalignment. Therefore, it is worthwhile to investigate the method of misaligning the mesh to obtain the ultimate performance of low IBF ratio.
\subsubsection{Toward large area detector}
However, the current results were tested with small area (25 mm$\times$25 mm) prototypes, it is a crucial issue to produce the DMM/TMM with proper sensitive area (i.e., 100 mm$\times$100 mm and larger), which is one of the biggest challenges for its application in a real experiment.
A dedicated fabrication process using thermal bonding method was designed and a larger DMM prototype with an active area of 160 mm$\times$160 mm was successfully fabricated.
\subsubsection{Studies on specific application}
The demand of real experiments is the greatest motivation for detector R\&D, our studies will focus on the potential applications of the high-rate TPCs and Gas-PMTs, where the absolute backflow ion number (gain $\times$ IBF) of $\sim$ 1 at a gain of higher than 1000 and low IBF ratio of 0.01\% at a high gain of 10$^5$ are expected individually.

\subsection{Summary}
We have developed and optimized a double micro-mesh gaseous structure (DMM) with two avalanche stages, which has a low IBF ratio and high gain. An IBF ratio as low as 0.025\% was obtained and a gain of up to 3$\times 10^6$ was reached for single electrons for the DMM. A triple micro-mesh gaseous structure (TMM) was also developed to further explore the potential of the multi-mesh method, excellent high gain and ultra-low IBF of $\sim$ 0.003\% is achieved, which is the lowest IBF value ever obtained in gaseous detectors. These features of the DMM and TMM present their strong potential for Gas-PMTs, RICH photoelectric readout, and high-rate TPCs.







\end{document}